\documentclass[preprintnumbers,showpacs,amsmath,amssymb,floatfix,9pt,prd,onecolumn,
superscriptaddress,nofootinbib]{revtex4}
\usepackage{CJK}
\usepackage{graphicx}
\usepackage{latexsym}
\usepackage{epsfig}
\usepackage{enumerate}
\usepackage{amssymb}
\usepackage[utf8]{inputenc}
\begin{document}
\title{Stable and self consistent charged gravastar model within the framework of $f(R,\,T)$ gravity }

\author{Piyali Bhar \footnote{Corresponding author}}
\email{piyalibhar90@gmail.com , piyalibhar@associates.iucaa.in}
\affiliation{Department of
Mathematics,Government General Degree College, Singur, Hooghly, West Bengal 712 409,
India}

\author{Pramit Rej}
\email{pramitrej@gmail.com
 } \affiliation{Department of
Mathematics, Sarat Centenary College, Dhaniakhali, Hooghly, West Bengal 712 302, India}

\begin{abstract}

In this work, we discuss the configuration of a gravastar (gravitational vacuum stars) in the context of $f(R, \,T )$ gravity by employing the Mazur-Mottola conjecture [P. Mazur and E. Mottola, Report No. LA-UR-01-5067; P. Mazur and E. Mottola, Proc. Natl. Acad. Sci.
USA $101$, $9545$ ($2004$)]. Gravastar is conceptually a substitute for a black hole theory as available in literature and it has three regions with different equation of states. By assuming that the gravastar geometry admits conformal killing vector, the Einstein-Maxwell field equations have been solved in different regions of gravastar by taking a specific equation of state as proposed by Mazur and Mottola. We match our interior spacetime to the exterior spherical region which is completely vacuum and described by Reissner-Nordstr\"{o}m geometry. For a particular choice of $f(R,\,T)$ as $f(R, \,T )=R+2\gamma T$, here we analyze various physical properties of the thin shell and also presented our results graphically for these properties. The stability analysis of our present model is also studied by introducing a new parameter $\eta$ and we explored the stability regions. Our proposed gravastar model in presence of charge might be treated as a successful stable alternative of the charged black hole in the context of this gravity.

\end{abstract}

\maketitle

\section{Introduction}

Over past few years, we have witnessed a considerable growing interest to study {\em gravastars} \cite{grav1,grav2,grav3,grav4,grav5,grav6,grav7,grav8} (and
further references therein), the gravitational vacuum star as it was proposed as an alternative theory of black holes. In $2001$, Mazur and Mottola (MM) \cite{mazur2001} first proposed a new idea for gravastars (collapsing stellar object) by extending the Bose-Einstein condensate (BEC) theory in the gravitating system. They further developed the theory in $2004$ \cite{mazur2004}. This MM model gives us an stable idea about the endpoint of gravitational collapse in the form of cold, dark, compact objects having mass above some critical values and provides solution to the classical black hole problems. After the pioneer innovation of Gravitational wave (GW) in 2015 \cite{abbott2016}, it is assumed that the GWs arise due to the merging of two massive black holes. But no observational proof for this theory. In this situation, the gravastar may play a crucial role to describe the final stage of the stellar evolution. Instead of no sufficient observational evidences in favour of gravastars directly for their existence, it is so much important to study the concept of gravastar that can be claimed as a feasible alternative to understand the concept of the black holes (BH).\par
The proposed model \cite{mazur2004} is a static spherically symmetric perfect fluid model having three different regions designated by: (I) interior region ($0\leq r_1<r$), (II) thin Shell region ($r_1< r< r_2$), (III) exterior region $(r_2< r)$ and it is separated by a thin shell of stiff
matter. In the interior region of the gravastar the relation between pressure and density is given by $p=-\rho$, inside the thin shell it is described by $p=\rho$ and finally in region III $p=\rho=0$. Where $p$ represents the isotropic pressure, $\rho$ is the matter density of the perfect fluid sphere and $r_2-r_1=\epsilon$ is the thickness of the shell, where $\epsilon \ll 1 $, because in a gravastar the thickness is very very small compared to its size. For an uncharged model of gravastar in ($3+1$)-D, the exterior spacetime is described by Schwarzschild geometry \cite{Schwarzschild1916}, whereas in case of charged gravastar model, the exterior spacetime is described by Reissner-Nordstr\"{o}m geometry \cite{reissner1916,nordstrom1918}.

The idea of gravastars has been discussed several times in the literature as an alternative to BH theory based on different mathematical as well as physical aspects. But, most of the investigations have been carried out by several workers in the framework of Einstein's general relativity (EGR) \cite{visser2004,cattoen2005,carter2005,bilic2006,debenedictis2006,horvat2007,rocha2008,horvat2008,turimov2009,usmani2011,lobo2013,
bhar2014,rahaman2015} (and further references therein). Though, it is well known that EGR is very well equipped to unveil many hidden mysteries behind nature, but this theory fails to explain the phenomenon of expanding universe along with the existence of dark matter \cite{riess1998,perlmutter1999,de2000,peebles2003,padmanabhan2003,clifton2012}. These two are the most important aspects of modern cosmology that have been accepted on the background of observational data. It is examined that Einstein theory of gravitation breaks down
at large scales, and a more generalized form of action is
required to describe the gravitational field at large scales. Therefore, the idea of coupling between matter and curvature produces several alternative modified theories to overcome the situation such as, $f(R)$ gravity \cite{de2010,capozziello2008}, $f(\mathbb{T})$ gravity ($\mathbb{T}$ is the torsion) \cite{ferraro2007,linder2010}, $f(R,\,T)$ gravity \cite{Harko:2011kv}, $f(R, T, R_{\gamma \eta}, T^{\gamma \eta})$ theory \cite{haghani2013,odintsov2013} and $f(\mathcal{G}, T )$ gravity \cite{sharif2016}, where $R$ indicates the Ricci scalar, T denotes the trace of the stress-energy tensor (SET) and $\mathcal{G}$ indicates the Gauss-Bonnet invariant.

Among all these theories, $f(R, \,T )$ theory has gained so much importance to describe various astrophysical stellar objects corresponding to different formulations \cite{sharif2017,das2017a,sharif2018aa,deb2018,sharif2018ab,sharif2018ac,sharif2020}. In this theory, gravitational Lagrangian is given by an arbitrary function of $R$ and $T$. Note that such dependence on $T$ may come due to exotic imperfect fluid or by considering quantum effects (case of conformal anomaly). Harko et al. \cite{Harko:2011kv} were the pioneers who first presented the formulation of $f(R, \,T )$ gravity. The rapid grow of attention on gravastar has motivated the researchers and scientists to discuss the outcomes of modified gravity theories on physical properties of gravastar. Das et al. \cite{das2017a} investigated the speculation of gravastar and studied its features by graphically with respect to different EoS in the $f(R, \,T )$ gravity framework. There are several applications in literature of $f(R, \,T )$ gravity theory to different cosmological domain \cite{moraes2014,moraes2015,moraes2016,singh2014,rudra2015,shabani2013,shabani2014,reddy2013,kumar2015,shamir2015,fayaz2016}. Among several applications, it is worthy to mention the references \cite{sharif2014,noureen2015,noureen2015a,zubair2015,noureen2015b,zubair2016,alhamzawi2016,moraes2016a,yousaf1,yousaf2,
maurya2020,buchdahl1959}. Sharif and Yousaf \cite{sharif2014} have studied the factors which affects the stability of a locally isotropic spherical self-gravitating systems within $f(R, \,T )$ gravity. A perturbation scheme has been employed on dynamical equations to find the collapse equation by  Noureen and Zubair \cite{noureen2015} and the condition on adiabatic index $\Gamma$ is constructed for Newtonian and post-Newtonian eras to address instability problem. Again, in their later work \cite{noureen2015a}, they presented a dynamical analysis of a spherically symmetric collapsing star under $f(R, \,T )$ gravity for an anisotropic environment with zero expansion. Zubair and Noureen \cite{zubair2015} then analyzed the gravitating sources carrying axial symmetry in $f(R, \,T )$ gravity. Also, the implications of the shear-free condition on the instability range of an anisotropic fluid has been investigated in $f(R, \,T )$ gravity by Noureen et al. \cite{noureen2015b}. Zubair et al. \cite{zubair2016} reported about the investigations on the possible formation of compact stars in $f(R, \,T )$ gravity. Alhamzawi and Alhamzawi \cite{alhamzawi2016} derived a new type of solution for $f(R, \,T )$ gravity and conclude about the contribution to gravitational lensing by modified gravity. Moraes et al. \cite{moraes2016a} studied the hydrostatic equilibrium configuration of neutron stars and strange stars in $f(R, \,T )$ gravity. The evolutionary behaviors of compact objects in $f(R, \,T )$ gravitational theory have been investigated by Yousaf et al. \cite{yousaf1} using structural scalars whereas in other work \cite{yousaf2} they examined the irregularity factors for a self-gravitating spherical star evolving in the presence of imperfect fluid under same gravity. Maurya et al.\cite{maurya2020} studied the hydrostatic equilibrium of stellar objects in modified $f(R, \,T )$ gravity that do not conserve energy momentum using {\em Buchdahl ansatz} \cite{buchdahl1959}. There are many other related works on the modified $f(R, \,T )$ gravity theory upon different physical aspects.

The effect of charge on the model of compact star is always important. The analysis of Raychaudhuri on charged dust distributions showed that
the conditions for collapse and oscillation depend on the ratio of matter density to charge density \cite{raychaudhuri1975spherically}. The gravitational collapse of a fluid
sphere to a point singularity may be avoided in the presence of large amounts of electric
charge during an accretion process onto a compact object proposed by De Felice et al. \cite{de1995relativistic}. Varela investigated charged object with neutral core and the electric charge distributed
on a d-shell \cite{Varela:2006qa}. Ivanov \cite{Ivanov:2002jy} proposed that the presence of the charge function serves
as a safety valve, which absorbs much of the fine-tuning,
necessary in the uncharged case. Bonnor \cite{bonnor1960mass} estimated the contribution of the electric field energy to the gravitational mass using certain special models. Debnath \cite{Debnath:2019eor} showed that the charged has non-negligible effect on different physical quantities in the Rastall-Rainbow Gravity. Rahaman et al. \cite{Rahaman:2012wc} proposed the model of charged gravastar in ($2+1$) dimensional gravity in anti de-sitter spacetime. Bhatti et al. \cite{Yousaf:2016yyb} investigated the role of different fluid parameters particularly electromagnetic field and $f(R)$ corrections on the evolution of cylindrical compact object. Motivated by all of these previous work done, in our present article we want to check the effect of charged on gravastar model in $f(R,\,T)$ gravity.\par

Very recently, the authors have modeled a charged (3+1)-dimensional gravastar under $f(\mathbb{T})$ modified gravity admitting conformal motion \cite{bhar2021} within the formulation of MM model \cite{mazur2004}. In this present work, we make an attempt to study a charged gravastar in the background of conformal symmetry of the spacetime in $f(R,\,T)$ modified gravity. Particular emphasis was given to obtain different physical features of the stellar object and its significance in describing expanding universe. In fact our earlier performed investigations on stellar object under modified gravity\cite{bhar2017,bhar2019,bhar2020,bhar2020a,rej2021,rej2021a} inspired us to consider this alternative formalism to the case of the gravastar, the final stage of the stellar evolution. Here we present the graphical variations of different physical features of gravastar for $f(R,\,T )$ model.\par

We adopt the following set up for the presentation of our paper. Next section displays the fundamental formulation of this theory with Conformal symmetry. Section \ref{sec3} expresses three geometries of gravastar: Interior spacetime, Thin shell and Exterior spacetime. Several physical properties of our model, $\it viz.$ the EoS parameter, proper shell thickness, entropy, energy content have been discussed in Section \ref{sec4}. The stability of the model is presented in the next section. Some discussions on our work, possibilities of observationally detection of a gravastar and some conclusions are provided in Section \ref{sec5}.


\section{Einstein-Maxwell Equation in $f(R,\,T)$ Gravity and Conformal Symmetry}\label{sec2}
In $(3+1)$-dimension, the interior of a static spherically symmetry spacetime is described by the following line element,
\begin{equation}\label{line}
ds^{2}=e^{\nu(r)}dt^{2}-e^{\lambda(r)}dr^{2}-r^{2}(d\theta^2+\sin^2\theta d\phi^2).
\end{equation}
where $\nu$ and $\lambda$ are two unknown functions of the radial co-ordinate `r' and independent on time, i.e., the metric coefficients are static. In our present discussion we use the gravitational or geometricized unit i.e., $G=1=c$.  For asymptotically flat
spacetime both the metric potential $\nu(r)$ and $\lambda(r)$ tends to $0$ as $r~\rightarrow~\infty$. For our present paper we have taken the signature of the spacetime as $(+,\, -,\, -,\,-)$.\\
In the presence of charge, the action in $f(R,\,T)$ theory of gravity is given as
\cite{Harko:2011kv},
\begin{eqnarray}\label{action}
S&=&\int \left[\frac{1}{16 \pi} f(R,T)+ \mathcal{L}_m+ \mathcal{L}_e\right]\sqrt{-g} d^4 x,
\end{eqnarray}
where $g = det(g_{ij}$), $f(R,\,T )$ represents the general function of Ricci scalar $R$ and trace $T$ of the energy-momentum tensor $T_{\mu \nu}$, $\mathcal{L}_m$ and $\mathcal{L}_e$ respectively denote the matter Lagrangian and Lagrangian for the electromagnetic field. Varying the action (\ref{action}) with respect to the metric $g_{\mu\nu}$, the field equations in $f(R,\,T)$ gravity can be obtained as \cite{Deb:2018gzt},
\begin{eqnarray}\label{gij}
G_{ij}&=&\frac{1}{f_R}\left[8\pi T_{ij}+\frac{1}{2}f g_{ij}-\frac{1}{2}Rf_Rg_{ij}-f_T (T_{ij}+ \Theta_{ij})\right.\nonumber\\&&\left.-(g_{ij}\Box-\nabla_{i}\nabla_{j})f_R+8\pi E_{ij}\right],
\end{eqnarray}
Where, $f=f(R,T)$, $f_R(R,T)=\frac{\partial f(R,T)}{\partial R},~f_T(R,T)=\frac{\partial f(R,T)}{\partial T}$. $\nabla_{\nu}$ represents the covariant derivative associated with the Levi-Civita connection of $g_{ij}$, $\Theta_{ij}=g^{\alpha \beta}\frac{\delta T_{\alpha \beta}}{\delta g^{ij}}$ and
$\Box \equiv \frac{1}{\sqrt{-g}}\partial_{i}(\sqrt{-g}g^{ij}\partial_{j})$ represents the D'Alembert operator, $T_{ij}$ is the energy momentum tensor given by,
\begin{eqnarray}\label{tmu1}
T_{ij}&=&-\frac{2}{\sqrt{-g}}\frac{\delta \sqrt{-g}\mathcal{L}_m}{\delta \sqrt{g_{ij}}},
\end{eqnarray}
Assuming that the matter Lagrangian $\mathcal{L}_m$ rely solely on $g_{ij}$ so that we obtain,
\begin{eqnarray}
T_{ij}&=& g_{ij}\mathcal{L}_m-2\frac{\partial \mathcal{L}_m}{\partial g^{ij}},
\end{eqnarray}
Now, the matter Lagrangian density $\mathcal{L}_m$ could be a function of pressure or density or both density and pressure. For our present paper, we choose the matter Lagrangian as $\mathcal{L}_m=\rho$ and the expression of $\Theta_{ij}=-2T_{ij}+\rho g_{ij}$, where $\rho$ is the matter density in modified gravity. This particular choice of Lagrangian matter density is based upon the pioneer work of Harko et al. \cite{Harko:2011kv}. In their work, they presented the field equations of several particular models, corresponding to some explicit forms of the function
$f(R,\,T)$. Faraoni \cite{Faroni2009} revisited the issue of the correct Lagrangian description of a perfect fluid $\mathcal{L}_m = p$ versus
$\mathcal{L}_m = -\rho$ in relation with modified gravity theories in which galactic luminous matter couples
nonminimally to the Ricci scalar and concluded that Lagrangians are only equivalent when the fluid couples
minimally to gravity and not otherwise. Bhar \cite{bhar2020a} presented spherically symmetric isotropic strange star
model under the framework of $f(R,\,T)$ theory of gravity by assuming $\mathcal{L}_m=\rho$. For our present model the energy-momentum tensor is given by,
\begin{eqnarray}
T_{ij}&=& (\rho+p)\chi_{i}\chi_{j}-p g_{ij},
\end{eqnarray}
where $\chi^{i}$ is the fluid four velocity satisfying $ \chi^{i}\chi_{j}=1$, $p$ is the isotropic pressure in modified gravity.\\
Again, $\mathcal{L}_e$  in eq. (\ref{action}) representing Lagrangian of the electromagnetic field is defined as, \[\mathcal{L}_e=-\frac{1}{16\pi}F_{\alpha \beta}F_{\gamma \delta} g^{\alpha \gamma}g^{\beta \delta}\]

where, $F_{ij}$ is the antisymmetric
electromagnetic field strength tensor defined by
\begin{eqnarray}
F_{ij}&=&\frac{\partial A_{j}}{\partial x^{i}}-\frac{\partial A_{i}}{\partial x^{j}},
\end{eqnarray}
and it satisfies the Maxwell equations,
\begin{eqnarray}
F^{ij}_{;j}=\frac{1}{\sqrt{-g}}\frac{\partial}{\partial x^{j}}(\sqrt{-g}F^{ij})&=&-4\pi \mathcal{J}^{i},\label{ta}\\
F_{ij;\lambda}+F_{j \lambda;i}+F_{\lambda i;j}&=&0.
\end{eqnarray}
where $A_{j}=(\phi(r),\,0,\,0,\,0)$ is the four-potential and  $\mathcal{J}^{i}$ is the
four-current vector, defined by
\begin{eqnarray}
\mathcal{J}^{i}&=&\frac{\rho_e}{\sqrt{g_{00}}}\frac{dx^{i}}{dx^0},
\end{eqnarray}
where $\rho_e$ denotes the proper charge density.
The expression for the electric field can be obtained from Eq.~(\ref{ta}) as follows,
\begin{eqnarray}
F^{01}&=&-e^{\frac{\lambda+\nu}{2}}\frac{q(r)}{r^2},
\end{eqnarray}
The electromagnetic energy-momentum tensor $E_{ij}$ has the following form :
\begin{eqnarray}
E_{ij}&=&\frac{1}{4 \pi}\left(F_{i}^{\alpha}F_{j \alpha}-\frac{1}{4}F^{\alpha \beta}F_{\alpha \beta} g_{ij}\right),
\end{eqnarray}
Let, $q(r)$ represents the net charge inside a sphere of radius `r' and it can be obtained as,
\begin{eqnarray}
q(r)&=& 4\pi \int_0^r \rho_e e^{\frac{\lambda}{2}} r^2 dr.
\end{eqnarray}
Taking the covariant divergence of eq. (\ref{gij}), we get \cite{Harko:2011kv,Koivisto:2005yk,BarrientosO.:2014ska},
\begin{eqnarray}\label{conservation}
\nabla^{i}T_{ij}&=&\frac{f_T(R,T)}{8\pi-f_T(R,T)}\Big[(T_{ij}+\Theta_{ij})\nabla^{i}\ln f_T(R,T)\nonumber\\&&+\nabla^{i}\Theta_{ij}-\frac{1}{2}g_{ij}\nabla^{i}T-\frac{8\pi}{f_T}\nabla^{i}E_{ij}\Big].
\end{eqnarray}
From eqn.(\ref{conservation}), it is clear that $\nabla^{i}T_{ij}\neq 0$ if $f_T(R,T)\neq 0$ and hence the system will not be conserved like Einstein gravity. The divergence of the matter energy-momentum tensor in this theory is non-zero whereas in GR and $f(R)$ it is zero.  For this reason this theory allows to break both weak and strong equivalence principles in $f (R,\,T )$ gravity. Also, it is possible to recover $f(R)$ gravity under the constraint $f(T)$ = 0.  According to the weak equivalence principle, ``All test particles in a given gravitational field will undergo the same acceleration, independent of their properties, including their rest mass". The equation of motion in this modified theory is based on those features of the particle that are thermodynamic in character, such as pressure, energy, density, etc. Furthermore, the strong equivalence principle asserts that, ``The gravitational motion of a small test body depends only on its initial position and velocity,
and not on its configuration " \cite{kopeikin2011}. This principle is similarly violated in $f (R,\,T )$  theory, resulting in non-geodesic motion of particles along world lines. In the context of quantum theory, the non-zero divergence of the effective energy-momentum tensor can be linked to the violation of energy conservation in the scattering phenomena. According to this theory, energy non-conservation can result in an energy flow between the four-dimensional spacetime and a compact extra-dimensional metric \cite{Lobato2018}. Also, the non-conservatively of the matter energy-momentum tensor is related to irreversible matter creation processes, in which there is an energy flow between the gravitational field and matter due to the geometry-matter coupling, with particles permanently added to the space-time \cite{Prigogine1986,Prigogine1988}. The creation of matter is accompanied by an irreversible energy flow from the gravitational field to the created matter constituents.

Now we are in a position to choose the $f (R,\,T )$ function. There are several theoretical models corresponding to different matter contributions for $f(R,\,T)$ gravity in order to discuss
the coupling effects of matter and curvature components. Harko et al. \cite{Harko:2011kv} choose three forms of $f(R,\,T)$ functions (i) $f(R,\,T)= R + 2f(T)$, (ii) $f (R,\,T) = f_1(R) + f_2(T)$, where
$f_1(R)$ and $f_2(T)$ are arbitrary functions of $R$ and $T$, respectively and (iii) $f (R,\,T) = f_1(R) + f_2(R)f_3(T)$, where $f_i, i = 1, 2, 3$ are arbitrary functions of the argument. For our present work, we consider second form proposed by Harko et al. \cite{Harko:2011kv}
with $f_1(R)=R$ and $ f_2(T)=2\gamma T$. So for our present case,
\begin{eqnarray}\label{e8}
f(R,\,T)&=& R+2 \gamma T,
\end{eqnarray}
where $\gamma$ is some small positive constant. Harko et al. \cite{Harko:2011kv} proposed that for $\gamma~\rightarrow~0$, the Eq. (\ref{e8}) produces the field equations in General Relativity. The term $2 \gamma T$ induces time-dependent coupling between curvature and
matter.\\
Substituting this particular form of $f(R,\,T)$ function in eq. (\ref{gij}) the
field equation for $f(R, T)$ gravity theory reads
\begin{eqnarray}
G_{ij}= 8\pi (T_{ij}^{\text{eff}}+E_{ij}),
\end{eqnarray}
where, \[T_{ij}^{\text{eff}}=T_{ij}\left(1+\frac{\gamma}{4\pi}\right)+\frac{\gamma}{8\pi}(T-2\rho)g_{ij}.\]
The generalized Tolman-Oppenheimer-Volkoff (TOV) equation for our
present model in $f(R,\,T)$ gravity can be obtained as,
\begin{eqnarray}
\frac{\gamma}{2\gamma+8\pi}(\rho'+3p')+\frac{8\pi}{8\pi+2\gamma}\frac{q}{4\pi r^4}\frac{dq}{dr}&=&\frac{\nu'}{2}(\rho+p)+\frac{dp}{dr}.\nonumber\\
\end{eqnarray}


The Einstein-Maxwell field equations in $f(R,\,T)$ gravity are given by,
\begin{eqnarray}
\kappa \rho^{\text{eff}}+\frac{q^2}{r^4}&=&\frac{\lambda'}{r}e^{-\lambda}+\frac{1}{r^{2}}(1-e^{-\lambda}),\label{f1}\\
\kappa p^{\text{eff}}-\frac{q^2}{r^4}&=& \frac{1}{r^{2}}(e^{-\lambda}-1)+\frac{\nu'}{r}e^{-\lambda},\label{f2} \\
\kappa p^{\text{eff}}+\frac{q^2}{r^4}&=&\frac{1}{4}e^{-\lambda}\left[2\nu''+\nu'^2-\lambda'\nu'+\frac{2}{r}(\nu'-\lambda')\right], \label{f3}\nonumber\\
\end{eqnarray}
with $\kappa=8\pi$. The quantity $q(r)$ actually determines the electric field as,
\begin{eqnarray}
E(r)&=&\frac{q(r)}{r^2}.
\end{eqnarray}
where $\rho^{\text{eff}}$, $p^{\text{eff}}$ are respectively the density and pressures in Einstein gravity. where
\begin{eqnarray}
\rho^{\text{eff}}&=& \rho+\frac{\gamma}{\kappa}( \rho-3p),\label{r1}\\
p^{\text{eff}}&=& p+\frac{\gamma}{\kappa}(\rho+5p),\label{r2}
\end{eqnarray}
by taking $p_r=p_t=p$ in ref.\cite{bhar2020a}.
Here $E(r)$ represents the electric field of the charged fluid sphere.

A familiar way to relate the geometry with matter is to use conformal
symmetry under conformal killing vectors (CKVs) described by,
\begin{equation}\label{con}
\mathcal{L}_\xi  g_{ik}=\xi_{i;k}+\xi_{k;i}=\psi  g_{ik},
\end{equation}
where $\mathcal{L}$ and $\psi$ respectively denote the Lie derivative operator and the conformal factor. The vector $\xi$ generates the conformal symmetry such that the metric $g$ is conformally mapped onto itself along $\xi$.
Neither $\xi$ nor $\psi$ need to be static even through one
consider a static metric \cite{boh1,boh2}. The underlying spacetime is asymptotically flat for $\psi = 0$ and in this case the Weyl tensor will also vanish. $\psi=$ constant and $\psi = \psi(x,t)$ respectively give homothetic vector and  conformal vectors.
Model of compact stars, wormholes and gravastars have been obtained earlier by several researchers in the realm of conformal symmetry. Mafa Takisa et al. \cite{mafa} investigated the effect of electric charge in
anisotropic compact stars with conformal symmetry.  Mak and Harko \cite{mak1} modelled quark stars with
conformal motions in general relativity. Bohmer et al. \cite{bohw} have studied the traversable
wormholes under the assumption of spherical symmetry and
the existence of a non-static conformal symmetry. Bhar et al. \cite{bhar16} studied the possibility of sustaining static and spherically symmetric traversable wormhole geometries admitting conformal motion in Einstein gravity. Mustafa et al. \cite{mus} explored the wormhole solutions in $f(\mathcal{G},\, T)$ gravity by assuming
two sorts of matter density profiles, which satisfy the Gaussian and Lorentzian
noncommutative distributions. More research works on conformal motion can be found in refs. \cite{a1,a11,a111,a2,a3,a4,a5}.\\
For the line element (\ref{line}), the conformal killing equations are written as,
\begin{eqnarray}
\xi^{1}\nu'=\psi,\,
\xi^{4}=C_1,\,
\xi^{1}=\frac{\psi r}{2},\,
\xi^{1}\lambda'+2\xi^{1},_1=\psi
\end{eqnarray}
~~~~~Where `prime' and `comma' stand for the derivative and partial derivative with respect to `r' and $C_1$ is a constant.\\
The above equations yield,
\begin{eqnarray}
e^{\nu}&=&C_2^{2}r^{2}, \label{eq11}\\
e^{\lambda}&=&\left(\frac{C_3}{\psi}\right)^{2},\\
\xi^{i}&=&C_1\delta_{4}^{i}+\left( \frac{\psi r}{2}\right)\delta_1^{i}.\label{eq13}
\end{eqnarray}
Where $C_2$ and $C_3$ are constants of integrations.\par

Using eqns. (\ref{eq11})-(\ref{eq13}), Einstein-Maxwell field equations (\ref{f1})-(\ref{f3}) become :
\begin{eqnarray}
\kappa \rho+\gamma(\rho-3p)+\frac{q^2}{r^4}&=&\frac{1}{r^2}\left[1-\frac{\psi^2}{C_3^2}\right]-\frac{2\psi\psi'}{rC_3^2},\label{j1} \\
\kappa p+\gamma(\rho+5p)-\frac{q^2}{r^4}&=&\frac{1}{r^2}\left[3\frac{\psi^2}{C_3^2}-1\right],\label{j2}\\
\kappa p+\gamma(\rho+5p)+\frac{q^2}{r^4}&=&\frac{\psi^2}{r^2C_3^2}+\frac{2\psi\psi'}{rC_3^2}.\label{j3}
  \end{eqnarray}

We want to solve the field equations (\ref{j1})-(\ref{j3}) in three different regions of the charged gravastar in the next section.

\section{The model of a Gravastar}\label{sec3}
Our present work explores the geometrical model of gravastar in  the context of  $f(R,\,T)$  gravity in presence of charge. The gravastar is a bubble like structure enclosed by thin-shell while the outer region is entirely a vacuum and the R-N \cite{reissner1916,nordstrom1918} spacetime. Three different regions of gravastar with the following specified EoS, i.e., (i) the inner region is governed by  $ p + \rho = 0$ for $0 \leq r \leq r_1$ (ii) for the intermediate thin-shell, the relation between pressure and density is given by $p = \rho $ for $r_1 \leq r \leq r_2$  and (iii) the exterior spacetime is described by $p =\rho = 0$ for $ r_2 \leq r$. The interior and the exterior radii of gravastar are $r_1$ and $r_2$ respectively and it is also assumed that the width of the shell is $r_2-r_1=\epsilon $, which is extremely small.
\subsection{The Interior Geometry}
To describe the interior geometry we have to solve the eqns. (\ref{j1})- (\ref{j3}) by using the EoS proposed in \cite{mazur2001,mazur2004}. To do that we add eqns. (\ref{j1}) and (\ref{j2}), which gives,
\begin{eqnarray}\label{s1}
(\rho+p)(\kappa+2\gamma)&=&\frac{2\psi^2}{r^2C_3^2}-\frac{2\psi\psi'}{rC_3^2},
\end{eqnarray}
Now to solve the equation (\ref{s1}) in the interior of the gravastar, we consider the following equation of state (EoS)
\begin{equation}\label{f4}
p=-\rho,
\end{equation}
proposed by Mazur and Mottola \cite{mazur2001,mazur2004} which manifests the dark energy EoS and it acts along the radially outward direction to oppose the
collapse. The above equation is a special case of the equation $p=\omega \rho$ with $\omega=-1$. $p=\omega \rho$ is known as dark energy EoS.
It is familiar that dark energy quintessence is a possible candidate
responsible for the late-time cosmic accelerated expansion and motivated by this concept, Chapline \cite{Chapline:2005ph} and Lobo \cite{Lobo:2005uf}
proposed a generalization of the gravastar model. Lobo \cite{Lobo:2005uf} proposed that the notion of dark energy is that of a spatially homogeneous cosmic fluid, which can be
extended to inhomogeneous spherically symmetric spacetimes by considering the pressure
in the dark energy equation of state is a negative radial pressure.
Using the relationship between the matter density $\rho$ and isotropic pressure $p$ given in eqn.(\ref{f4}), from eqn. (\ref{s1}) we get the following ordinary differential equation which is linear in conformal factor $\psi$ as,
\begin{equation}\label{eq1}
\frac{2\psi}{r^2C_3^2}(\psi-\psi'r)=0,
\end{equation}
which gives two solutions for $\psi$, $\text{either}~~\psi=0 ~~\text{or}~~ \psi=A_1r.$
where $A_1$ is the constant of integration. Since $\psi=0$ implies the asymptotically flat spacetime, we take $\psi=A_1r$ to calculate the matter density, pressure and the other physical quantities. Invoking the expression of the conformal factor $\psi=A_1r$, the expressions for the metric coefficients, $\rho,\,p,\,E^2$ and $\rho_e$ are obtained as,
\begin{eqnarray}
e^{-\lambda}&=&A^2 r^2,\,
e^{\nu}=C_2^2r^2,\\
\rho&=&\frac{1 - 6 A^2 r^2}{2 (4 \gamma + \kappa) r^2}=-p,\label{s2}\\
E^2&=& \frac{1}{2r^2}\label{s3},\,
\rho_e(r)=\frac{A}{4\sqrt{2}\pi r}.
\end{eqnarray}
Here we have used the notation $A=\frac{A_1}{C_3}$ which is an another constant. The metric coefficients $e^{\lambda}$ is inversely proportional to $r^2$ but $e^{\nu}$ is directly proportional to $r^2$. Since in the interior region of the gravastar, the energy density is positive, from eqn. (\ref{s2}) we get $A^2<\frac{1}{6r^2}$ and it gives the upper bound for $A^2$. From eqn. (\ref{s3}), we get $A>0$. One can note that both the pressure and density are inversely proportional to $r^{2}$ and all pressure, density, electric field and charge density suffer from central singularities, i.e., for $r\rightarrow 0$, they blow up without bound at the center of charged gravastar and it is a natural behavior of the CKV model. The electric field $E^2$ is inversely proportional to $r^2$ and it does not depend on $A$. The charged density $\rho_e$ depends on $A$ and inversely proportional to $r$.
The active gravitational mass $M(r)$ can be obtained from the following formula,
\begin{eqnarray}
M(r)&=&4\pi\int_0^{r}\eta^{2}\left[\rho(\eta)+\frac{E^{2}(\eta)}{\kappa}\right]d\eta\nonumber\\
&=&\frac{r}{4}\left[1+\frac{\kappa}{4 \gamma + \kappa}(1 - 2 A^2 r^2)\right].
\end{eqnarray}
The mass function $M(r)$ does not suffer from central singularity since as $r\rightarrow 0$, $M(r)\rightarrow 0$. One can note that the active gravitational mass function depends on both $A$ and the coupling constant $\gamma$. Using the bound for $A^2$, we get the lower bound for the active gravitational mass as, $M(r)>\frac{r}{12}\left(\frac{12\gamma+5\kappa}{4\gamma+\kappa}\right)$.

\subsection{The Intermediate Thin Shell}
In the shell of the gravastar, following the concept of Mazur \& Mottola \cite{mazur2001,mazur2004}, the relation between the pressure $p$ and the energy density $\rho$ is taken as,
\begin{equation}\label{e2}
p=\rho ,
\end{equation}
This EoS is a special case of barotropic EoS $p=\alpha \rho$ with $\alpha=1$. In general, where the pressure is
the only function of density, i.e., $P = P(\rho)$, and vice-versa, is called barotropic fluids.
They are considered as unrealistic but their simplicity has a pedagogical value in illustrating the several
approaches used to solve different systems and ``physically"
interesting scenarios \cite{Hernandez:2020pcn}. In this connection, we want to mention that Zel'dovich \cite{Zeldovich72} first conceived the idea of this kind of fluid in connection with cold
baryonic universe and it was described
as the stiff fluid. Staelens et al. \cite{Staelens:2019sza} studied the spherical collapse of an over-density of a barotropic fluid with linear equation of state in a cosmological background. Bergh and Slobodeanu \cite{VandenBergh:2016djv} studied shear-free perfect fluids with a barotropic equation of state in general relativity. Rahaman along with his collaborators \cite{fr1} used the barotropic EoS to obtain a new class of exact solutions for the interior in $(2 + 1)$-dimensional spacetime by assuming isotropic pressure both with and without cosmological constant $\Lambda$. Wesson \cite{wesson1978exact} obtained spherically-symmetric and non static solution with an inhomogeneous density profile $\rho$ and a pressure $p$ given by the stiff equation of state $p=\rho c^2$, $c$ being a constant. Stiff fluid model has been used earlier by several researchers in the field of astrophysics as well as in cosmology that can be found in the refs. \cite{carr1975primordial,madsen1992evolution,braje2002rx,Ferrari:2007zzb}  \\
From eqns. (\ref{f1}) and (\ref{f2}) with the help of eqn. (\ref{e2}) we get the following ordinary differential equation (ODE),
\begin{equation}
\frac{2r\psi\psi'}{C_3^2}-\left(\frac{2\gamma-\kappa}{4\gamma+\kappa}\right)\frac{\psi^2}{C_3^2}=\frac{2\gamma+\kappa}{8\gamma+2\kappa},
\end{equation}
The above ODE is linear equation in $\psi^2$, which on integrating gives,
\begin{eqnarray}
  \frac{\psi^{2}}{C_3^2} &=& \frac{2\gamma+\kappa}{2\kappa-4\gamma}-Dr^{\frac{2\gamma-\kappa}{4\gamma+\kappa}},
  \end{eqnarray}
  where $D$ is a positive constant of integration.\par
 Now the metric co-efficients for the thin shell can be obtained as,
  \begin{eqnarray}
   e^{-\lambda} &=& \frac{2\gamma+\kappa}{2\kappa-4\gamma}-Dr^{\frac{2\gamma-\kappa}{4\gamma+\kappa}},\\
  e^{\nu}&=&C_2^2r^2.
  \end{eqnarray}
  \begin{figure}[htbp]
        \includegraphics[scale=.5]{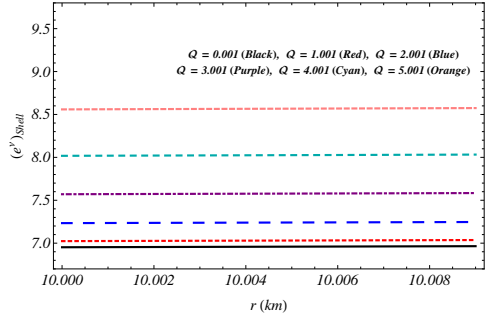}
       \caption{The metric coefficient $e^{\nu}$ inside the thin shell. \label{nu}}
\end{figure}

 \begin{figure}[htbp]
        \includegraphics[scale=.5]{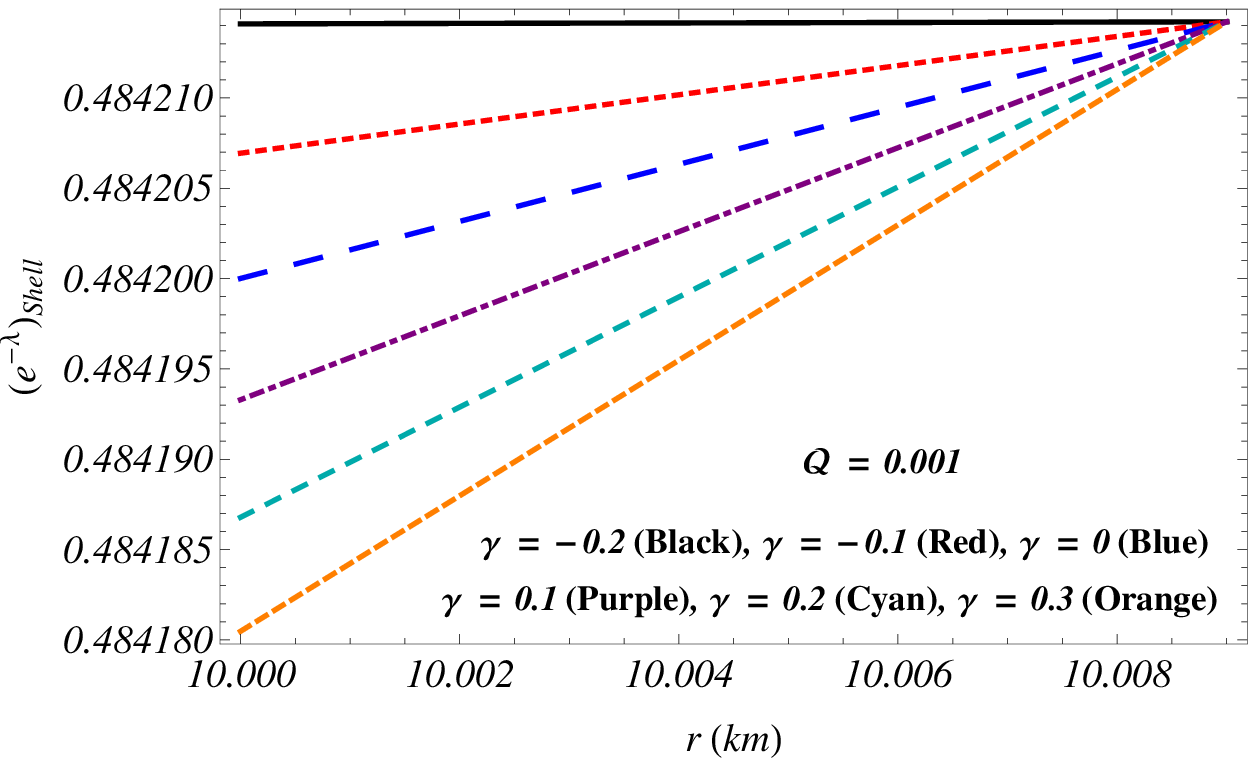}
         \includegraphics[scale=.5]{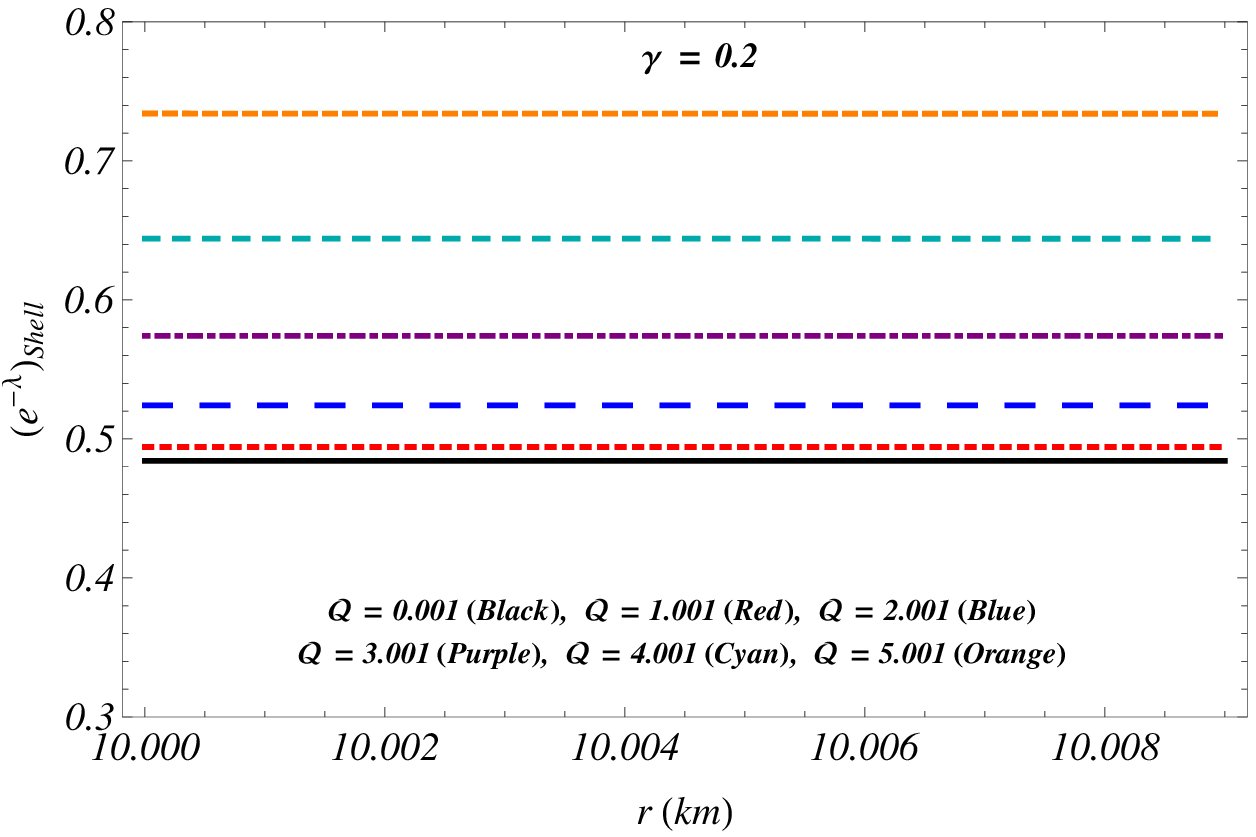}
       \caption{$e^{-\lambda}$ inside the thin shell. \label{lambda}}
\end{figure}
The profiles of the metric coefficients inside the thin shell are shown in Figs.~\ref{nu} and \ref{lambda}.\\
  The expressions for matter density and isotropic pressure inside the thin shell are obtained as,
  \begin{eqnarray}
  \rho &=& \frac{1}{2r^3}\Big[\frac{r}{\kappa-2 \gamma} - \frac{3 D r^{\frac{6 \gamma}{4 \gamma + \kappa}}}{
 4 \gamma + \kappa}\Big]= p ,
  \end{eqnarray}
  The electric field inside the thin shell takes the form
  \begin{eqnarray}
  E^2&=& \frac{2 \gamma}{(2 \gamma - \kappa) r^2} + \frac{
 3 D (2 \gamma + \kappa)}{
 2 (4 \gamma + \kappa)r^{\frac{6\gamma+3\kappa}{4\gamma+\kappa}}}, \label{e}
\end{eqnarray}
and the electric charged density $\rho_e$ can be obtained as,
\begin{widetext}
\begin{eqnarray}
\rho_e(r)&=& \frac{3 D (2 \gamma - \kappa) (2 \gamma + \kappa) (10 \gamma + \kappa) r^{\frac{
  2 \gamma}{4 \gamma + \kappa}} +
 8 \gamma (4 \gamma + \kappa)^2 r^{\frac{\kappa}{4 \gamma + \kappa}}}{8 (2 \gamma - \kappa) (4 \gamma + \kappa)^2 \pi r^{\frac{4(3\gamma+\kappa)}{4\gamma+\kappa}}\chi(r)}\Phi(r),\label{e}
\end{eqnarray}
\end{widetext}
where $\Phi(r)$ and $\chi(r)$ are functions of `r' and they depend on the coupling constant $\gamma$. Their expressions are given as,
\begin{eqnarray}
\Phi(r)=\sqrt{\frac{\kappa+2\gamma}{2(\kappa-2\gamma)}-Dr^{\frac{2\gamma-\kappa}{4\gamma+\kappa}}},\\
\chi(r)=\sqrt{\frac{8 \gamma}{(2 \gamma - \kappa) r^2} + \frac{
 6 D (2 \gamma + \kappa)}{
  (4 \gamma + \kappa)r^{\frac{6\gamma+3\kappa}{4\gamma+\kappa}}}}.
\end{eqnarray}
\begin{figure}[htbp]
        \includegraphics[scale=.5]{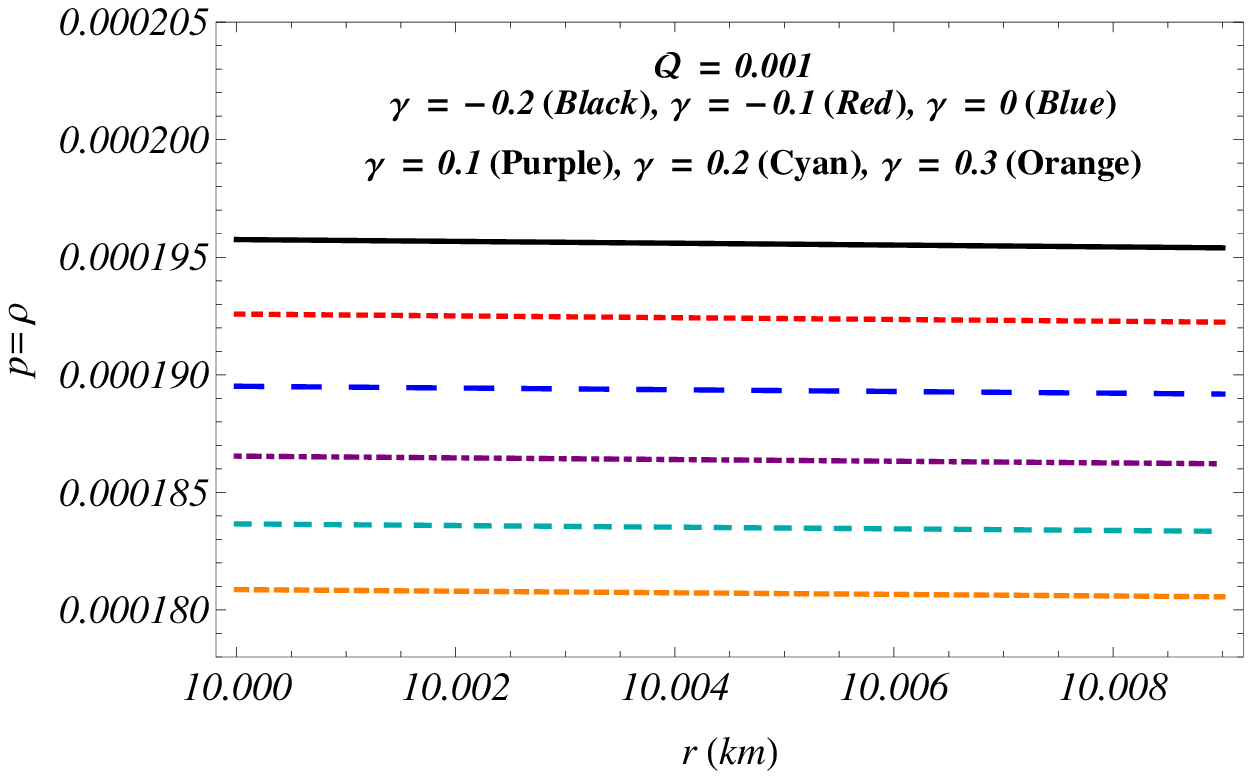}
        \includegraphics[scale=.5]{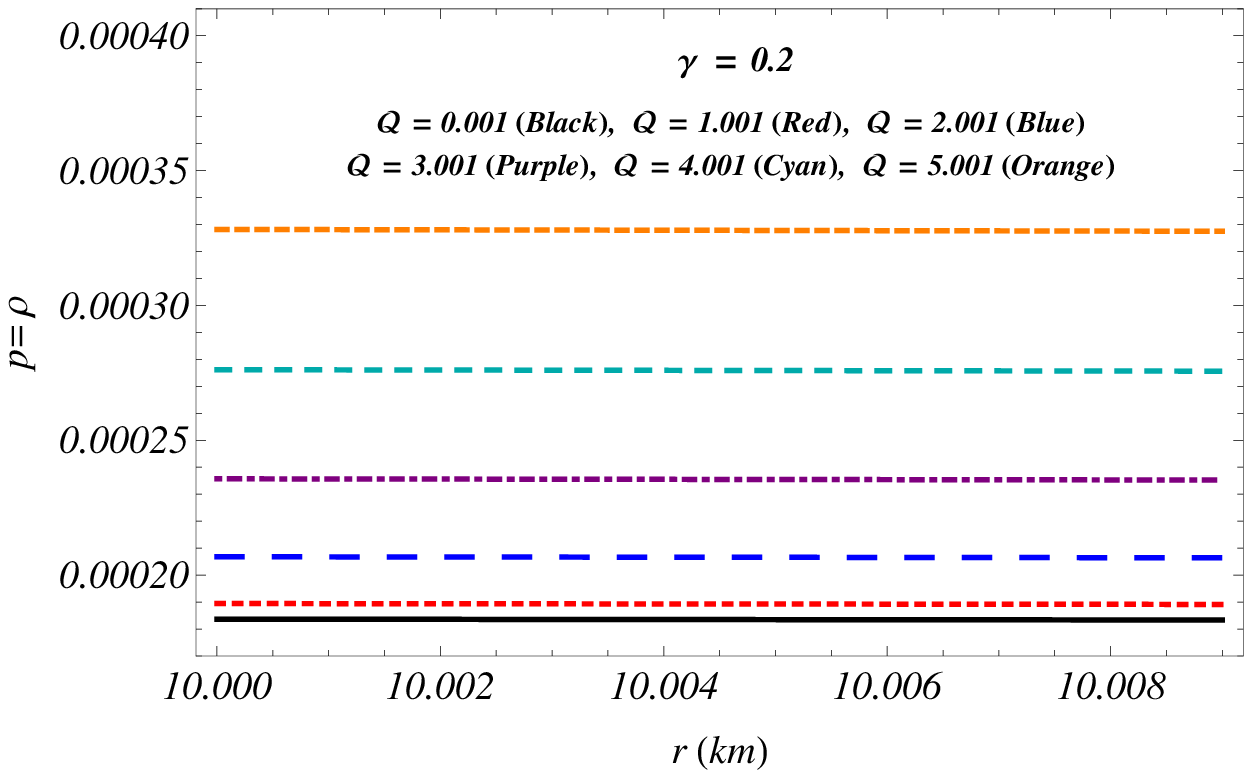}
       \caption{Variation of pressure and density inside the thin shell.\label{pr}}
\end{figure}
The profiles of pressure and density inside the thin shell are depicted in Fig.~\ref{pr}.

\subsection{Exterior Spacetime and junction condition}
For this region we consider, $ p=\rho=0 $ , which ensures that the exterior
space-time is described by the Reissner-Nordstr\"{o}m line element given by,
\begin{eqnarray}
ds^{2}&=& \mathcal{F} dt^{2}-\mathcal{F}^{-1}dr^{2}-r^{2}(d\theta^{2}+\sin^{2}\theta d\phi^{2}),\label{exterior}
\end{eqnarray}
where, $\mathcal{F}=\left(1-\frac{2\mathcal{M}}{r}+\frac{\mathcal{Q}^2}{r^{2}}\right)$, $\mathcal{M}$ being the mass and $\mathcal{Q}$ being the charge of the gravastar.
Instead of one junction surface that a compact star has, the gravastar configuration has two junction surfaces, since it is like a hollow sphere. One is between interior region and intermediate thin shell (i.e., at $r = r_1$) and the other is between the shell and exterior spacetime
(i.e., at $r = r_2$). Now for our present model of the gravastar, the metric potentials $g_{rr}$ and $g_{tt}$ must be continuous at the interface
between the core and the shell at $r = r_1$ (interior radius) and it gives the following relationship :
\begin{eqnarray}\label{m1}
A^2r_1^2&=&\frac{2\gamma+\kappa}{2\kappa-4\gamma}-Dr_1^{\frac{2\gamma-\kappa}{4\gamma+\kappa}},
\end{eqnarray}
Now using the matching condition between the shell and the
exterior region at $r = r_2$ (exterior radius) yields the following relationships:
\begin{eqnarray}
C_2^2r_2^2&=&1-\frac{2\mathcal{M}}{r_2}+\frac{\mathcal{Q}^2}{r_2^{2}},\label{m2}\\
\frac{2\gamma+\kappa}{2\kappa-4\gamma}-Dr_2^{\frac{2\gamma-\kappa}{4\gamma+\kappa}}&=&1-\frac{2\mathcal{M}}{r_2}+\frac{\mathcal{Q}^2}{r_2^{2}}.\label{m3}
\end{eqnarray}
Solving the eqns. (\ref{m1})-(\ref{m3}), we obtain the expressions for $C_2,\,D$ and $A$ as,
\begin{eqnarray}
C_2&=&\frac{1}{r_2}\sqrt{1-\frac{2\mathcal{M}}{r_2}+\frac{\mathcal{Q}^2}{r_2^{2}}},\\
D&=&r_2^{\frac{\kappa-2\gamma}{\kappa+4\gamma}}\left[\frac{2\mathcal{M}}{r_2}-\frac{\mathcal{Q}^2}{r_2^{2}}-\frac{\kappa-6\gamma}{2\kappa-4\gamma}\right],\\
A&=&\frac{1}{r_1}\sqrt{\left[\frac{2\gamma+\kappa}{2\kappa-4\gamma}-Dr_1^{\frac{2\gamma-\kappa}{4\gamma+\kappa}}\right]}.
\end{eqnarray}
To determine the values of these constants, we consider mass of the gravastar $M = 1.75M_{\odot}$, inner radius $r_1 = 10$ km and outer radius $r_2 = 10.009$ km, $\mathcal{Q}=0.001$ which provides the numeric values of $C_2,\,D$ and $A$ for different values of the coupling constant $\gamma$ presented in table~\ref{table1}.\\

\begin{table}[t]
\centering
\caption{The numerical values of the constants $C_2,\,A$ and $D$ have been presented for different values of the coupling constant $\gamma$ for the compact star with $M =1.75~M_{\odot},\, r_1 = 10$ km., $r_2=10.009$ km,  $\mathcal{Q} = 0.001$.}\label{table1}
\begin{tabular}{@{}ccccccccccccc@{}}
\hline
$\gamma$&&$C_2$&$A$&$D$\\
\hline
-0.2&& $0.0695229$&$0.0695855$& $0.00134$\\
-0.1&&  $0.0695229$& $0.0695850$ & $0.08351$ \\
0.0&& $0.0695229$ & $0.0695845$ & $0.15800$\\
0.1&& $0.0695229$ & $0.0695840$ & $0.22573$\\
0.2&& $0.0695229$& $0.0695835$ & $0.28753$\\
0.3&&$0.0695229$& $0.0695831$& $0.34411$\\
\hline
\end{tabular}
\end{table}

The extremely cold radiation fluid in the shell is confined to region II by the surface tensions at the timelike interfaces $r_1$ and $r_2$ \cite{mazur2001,mazur2004}. When we are matching our interior spacetime to the exterior R-N spacetime we should keep in mind that here considered a hollow sphere with inner radius $r_1$ and outer radius $r_2=r_1+\epsilon$, and $\epsilon\ll r_1$ and according to Mazur and Mottola \cite{mazur2001,mazur2004}, $\epsilon$ does not exceed plank length. At the boundary we match our interior region to the exterior line element. Obviously the metric coefficients are continuous at $r=a$, but it does not ensure that their derivatives are also continuous at the junction surface. In other words the affine connections may be discontinuous there.
 The surface stress energy tensor is given by Lanczos equations in the following form \cite{dm1}
\begin{equation}
\mathcal{S}^{i}_{j}=-\frac{1}{8\pi}(\kappa^{i}_j-\delta^{i}_j\kappa^{k}_k),
\end{equation}
where the Latin indexes running as $i,\, j = t,\,\theta,\,\phi$. The factor $\kappa_{ij}$ represents the discontinuity in the extrinsic curvature $K_{ij}$ with
$\kappa_{ij}=\left[K_{ij}\right]^{+}-\left[K_{ij}\right]^{-}$, the expressions for $K_{ij}^{\pm}$ is given as,
\begin{equation}
K_{ij}^{\pm}=-n_{\nu}^{\pm}\left(\partial_je_i^{\nu}+\Gamma^{\nu}_{\alpha\beta}e_i^{\alpha}e_j^{\beta}\right),
\end{equation}
where $e_i^{\alpha}=\frac{\partial x^{\alpha}}{\partial \xi^i}$, $ \xi^i$ represents the coordinate on the shell, $n_{\nu}^{\pm}$ are the unit normal vectors on the surface with $n^{\nu}n_{\nu}=1$, $\Gamma^{\nu}_{\alpha\beta}$ is the Christoffel symbols and ``$+$" and ``$-$" signs correspond to exterior i.e., Reissner-Nordstr\"{o}m spacetime and interior spacetime respectively.\par
\begin{figure}[htbp]
        \includegraphics[scale=.5]{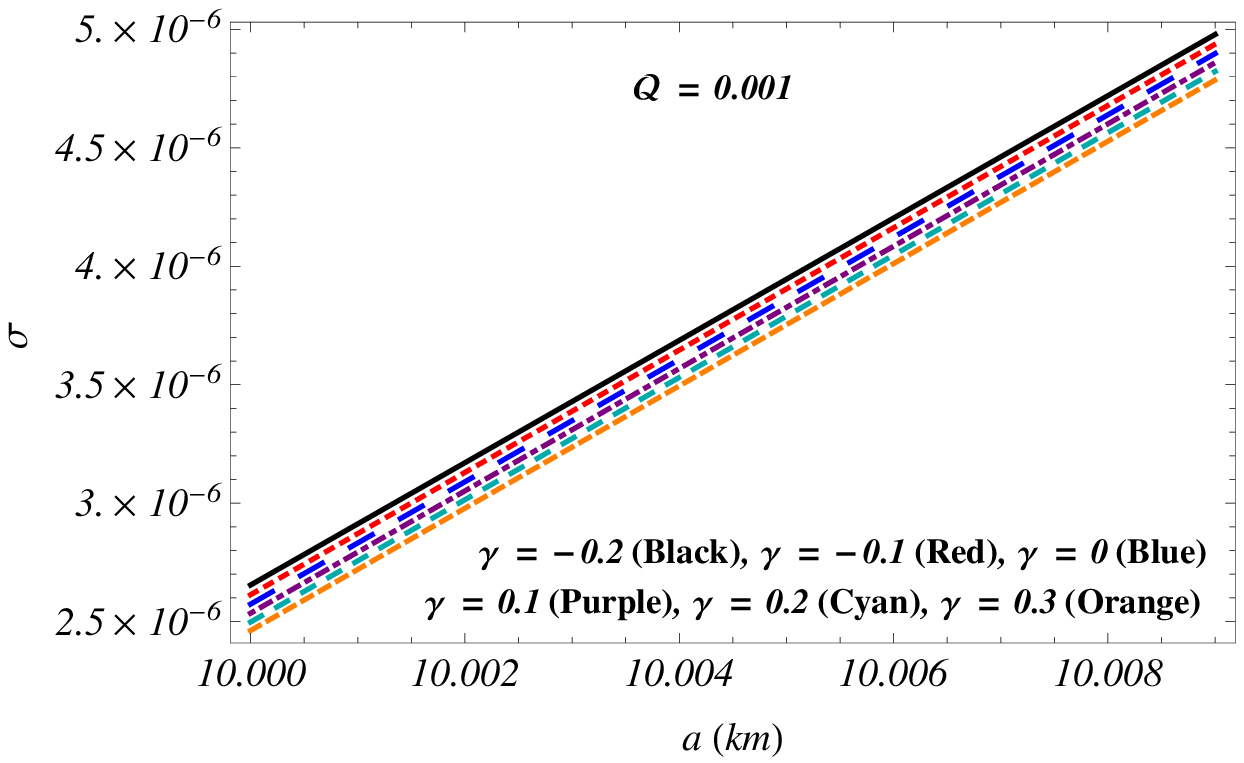}
         \includegraphics[scale=.5]{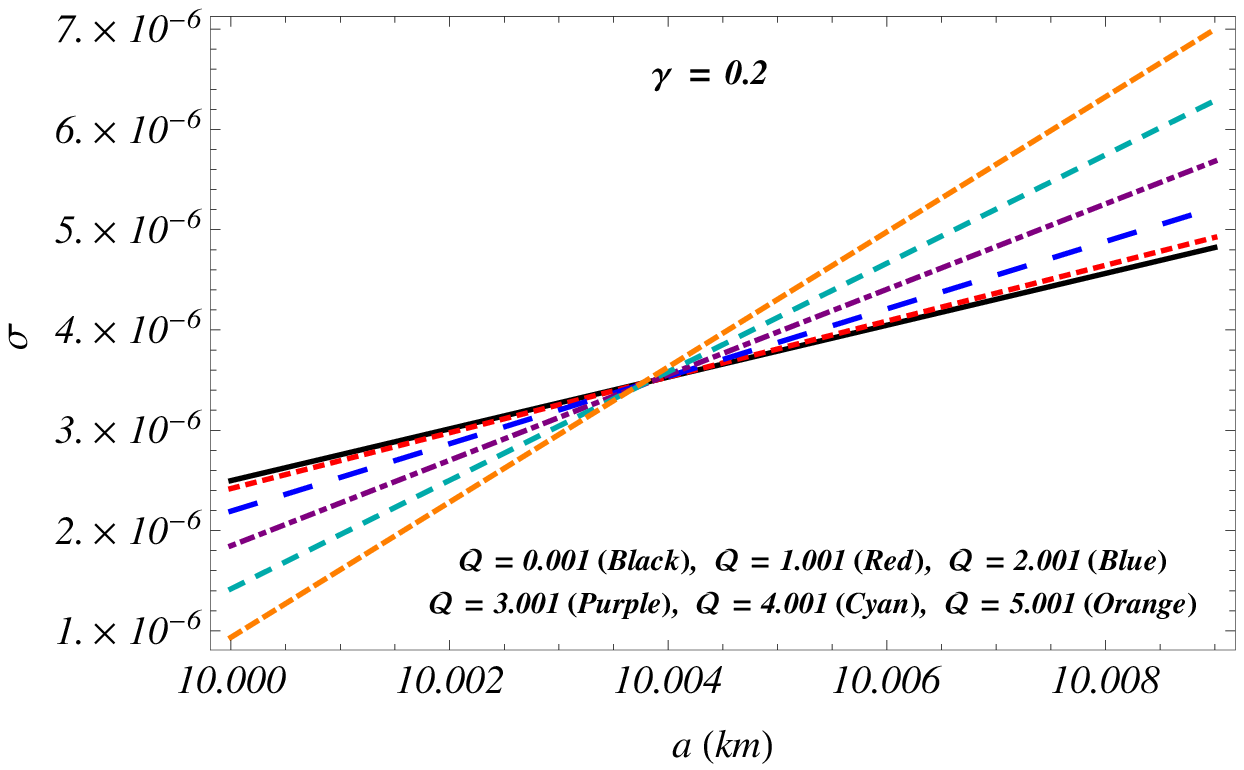}
       \caption{Variation of surface energy density inside the thin shell. \label{se}}
\end{figure}
Using the spherical symmetry nature of the spacetime, using Lanczos equation, the surface stress energy tensor can be written as $\mathcal{S}_{ij}=diag(\sigma,\,-\mathcal{P},\,-\mathcal{P},\,-\mathcal{P})$. Where $\sigma$ and $\mathcal{P}$ being the surface energy density and surface pressure respectively. The mathematical expressions for surface energy density $\sigma$ and the surface pressure $\mathcal{P}$ at the junction surface $r = a$ are obtained as \cite{das2017a},
\begin{eqnarray}
\sigma &=&-\frac{1}{4\pi a}\left[\sqrt{f} \right]_{-}^{+}=-\frac{1}{4\pi a}\left[\sqrt{1-\frac{2\mathcal{M}}{a}+\frac{\mathcal{Q}^2}{a^2}}-Aa\right],\label{sigma1}\nonumber\\
\\
\mathcal{P}&=&-\frac{\sigma}{2}+\frac{1}{16\pi}\left[\frac{f'}{\sqrt{f}}\right]_{-}^{+}\nonumber\\
&=&\frac{1}{8\pi a}\left[\frac{1-\frac{\mathcal{M}}{a}}{\sqrt{1-\frac{2\mathcal{M}}{a}+\frac{\mathcal{Q}^2}{a^2}}}-2Aa\right]\label{p1}.
\end{eqnarray}

\begin{figure}[htbp]
        \includegraphics[scale=.5]{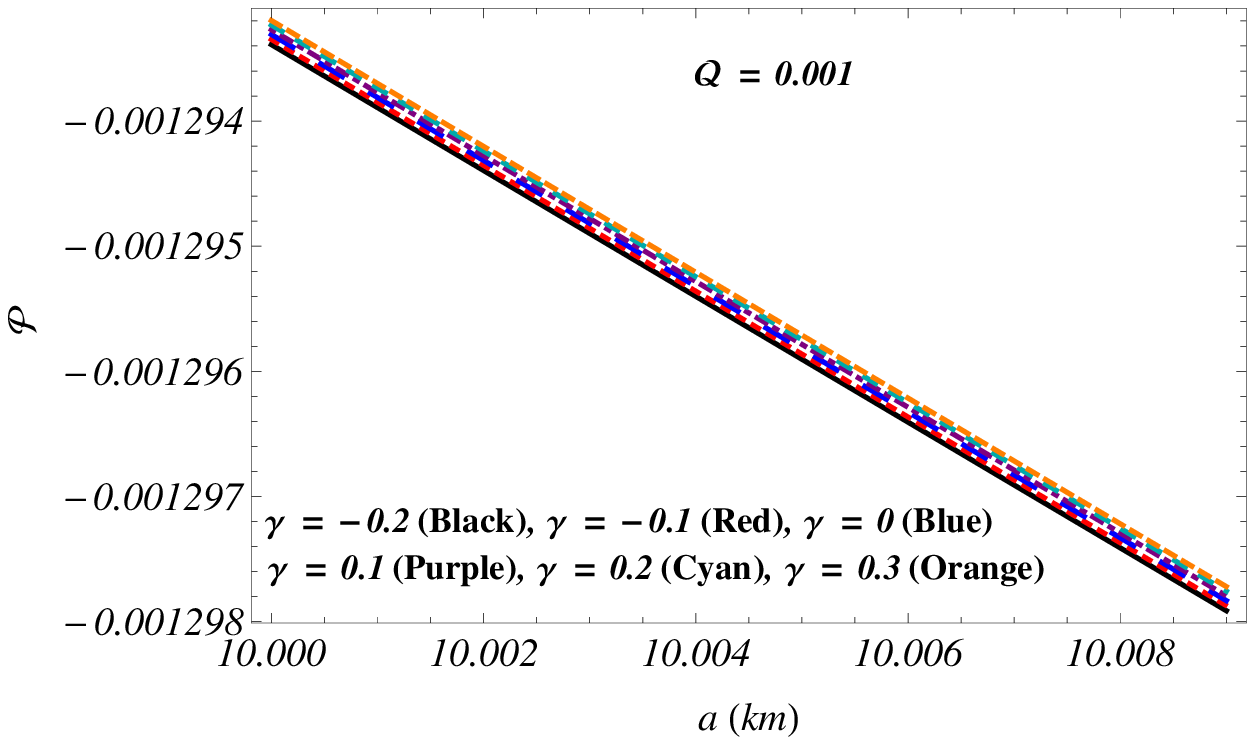}
         \includegraphics[scale=.5]{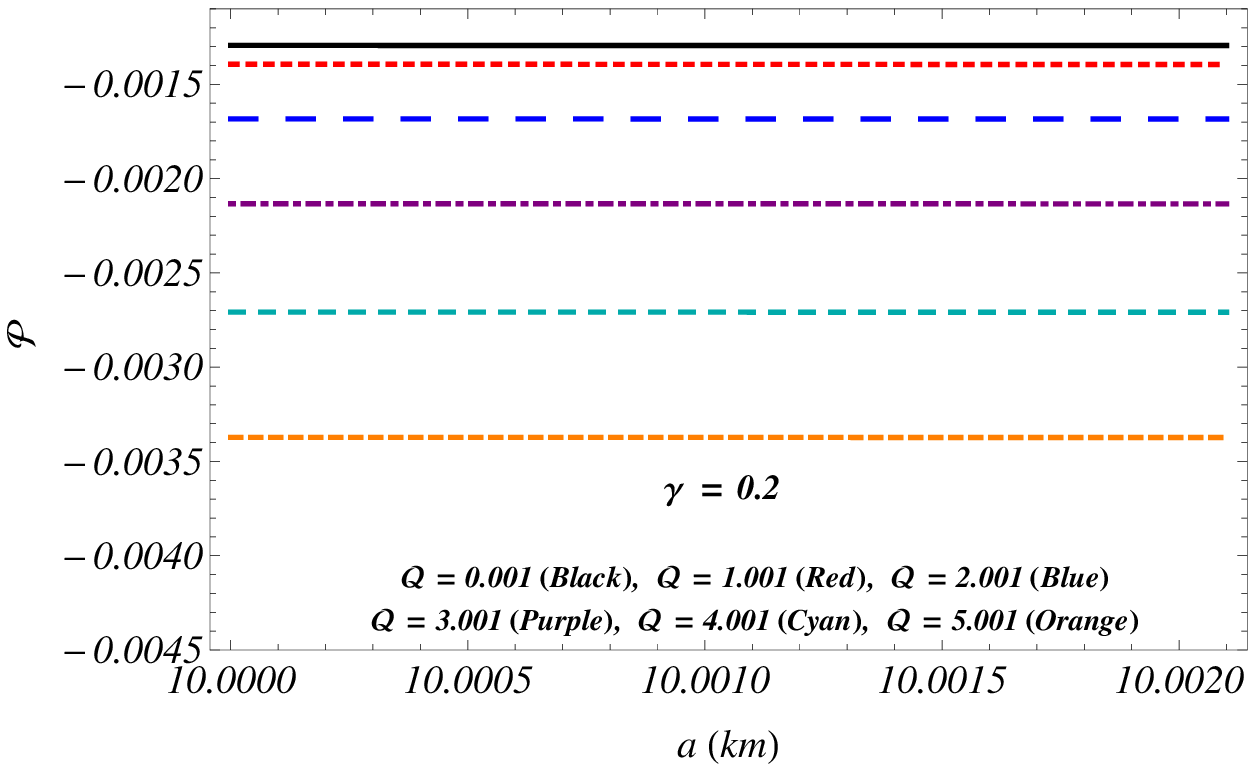}
       \caption{Variation of surface pressure inside the thin shell.\label{sp}}
\end{figure}
The profiles of surface energy density and surface pressure have been plotted in Fig.~\ref{se} and \ref{sp} respectively.\\
The mass of the thin shell ($m_{\text{shell}}$) of width $\epsilon$ can be obtained from the following formula:
\begin{eqnarray}\label{ms}
m_{\text{shell}}&=&4\pi a^2\sigma,\nonumber\\
&=&Aa^2-a\sqrt{1-\frac{2\mathcal{M}}{a}+\frac{\mathcal{Q}^2}{a^2}}.
\end{eqnarray}
Rearranging the above equation, the mass of the charged gravastar is calculated as,
\begin{eqnarray}
\mathcal{M}&=&\frac{a}{2}\left[1+\frac{\mathcal{Q}^2}{a^2}-\frac{m_{\text{shell}}^2}{a^2}-A^{2}a^2+2Am_{\text{shell}}\right],\\
&=&\frac{a^2+\mathcal{Q}^2}{2a}-\frac{a}{2}\left(\frac{m_{\text{shell}}}{a}-Aa\right)^2.\label{totalmass}
\end{eqnarray}
We see that the total mass $\mathcal{M}$ of the gravastar can not exceed $\frac{a^2+\mathcal{Q}^2}{2a}$. Moreover if the mass of the thin shell, radius of the gravastar and total charge $\mathcal{Q}$ are known, the total mass of the gravastar can be obtained from eqn. (\ref{totalmass}).

\section{Some physical properties of our present model}\label{sec4}
In this section we want to explore some physical features of the developed
structure, i.e., equation of state, proper length, entropy and energy contents
within the shell's region. Since the constructed geometry of gravastar is
the matching of two different spacetimes, the stiff perfect
fluid moves along these spacetimes through the shell region. The impact of electromagnetic field on different physical features of the charged gravastar in the context of $f(R,\,T)$ gravity will also be discussed.

\subsection{The EoS parameter}
The equation of state parameter $\omega$ for our present model is written as,
\begin{eqnarray}
\omega(a)=\frac{\mathcal{P}(a)}{\sigma(a)},
\end{eqnarray}
Now using the eqns. (\ref{sigma1}) and (\ref{p1}), we obtain the expression for $\omega(a)$ as,
\begin{eqnarray}\label{omega1}
\omega(a)=\frac{1}{2}\left[\frac{\frac{1-\frac{\mathcal{M}}{a}}{\sqrt{1-\frac{2\mathcal{M}}{a}+\frac{\mathcal{Q}^2}{a^2}}}-2Aa}{Aa-\sqrt{1-\frac{2\mathcal{M}}{a}+\frac{\mathcal{Q}^2}{a^2}}}\right].
\end{eqnarray}
\begin{figure}[htbp]
        \includegraphics[scale=.5]{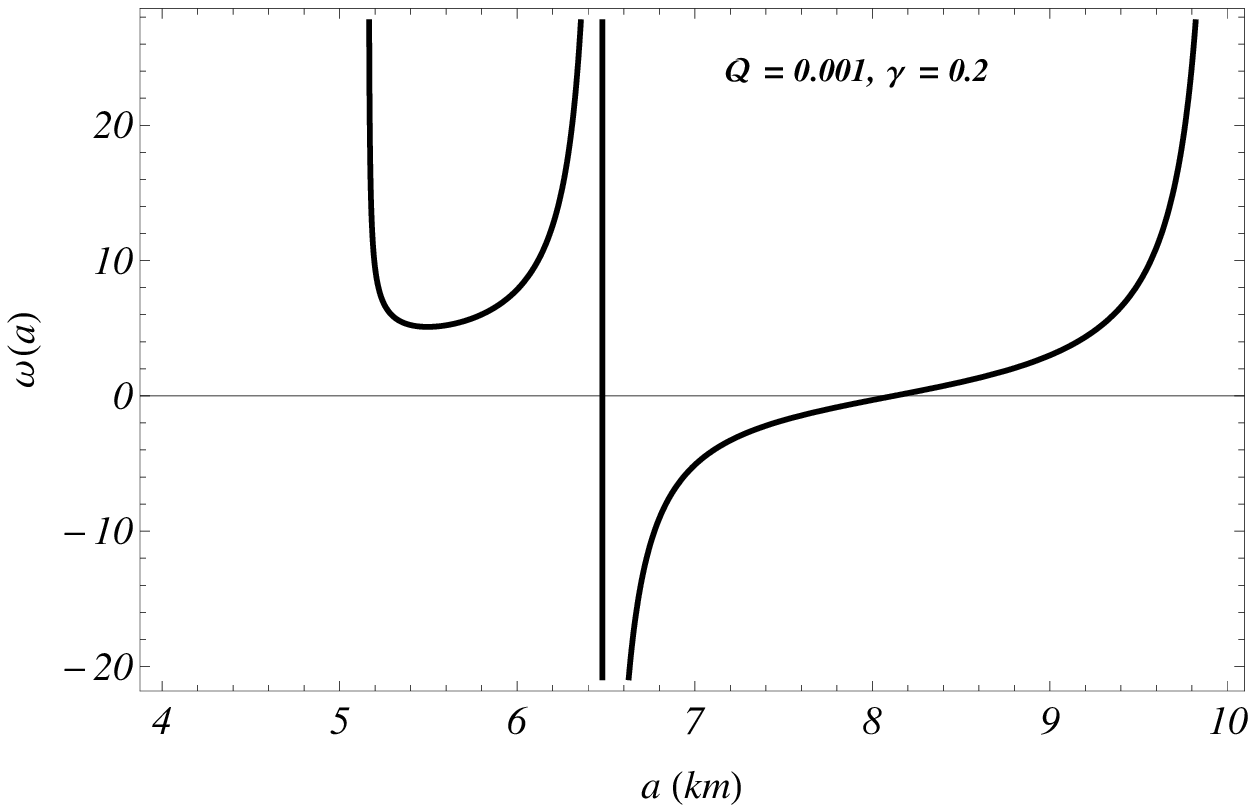}
        \includegraphics[scale=.5]{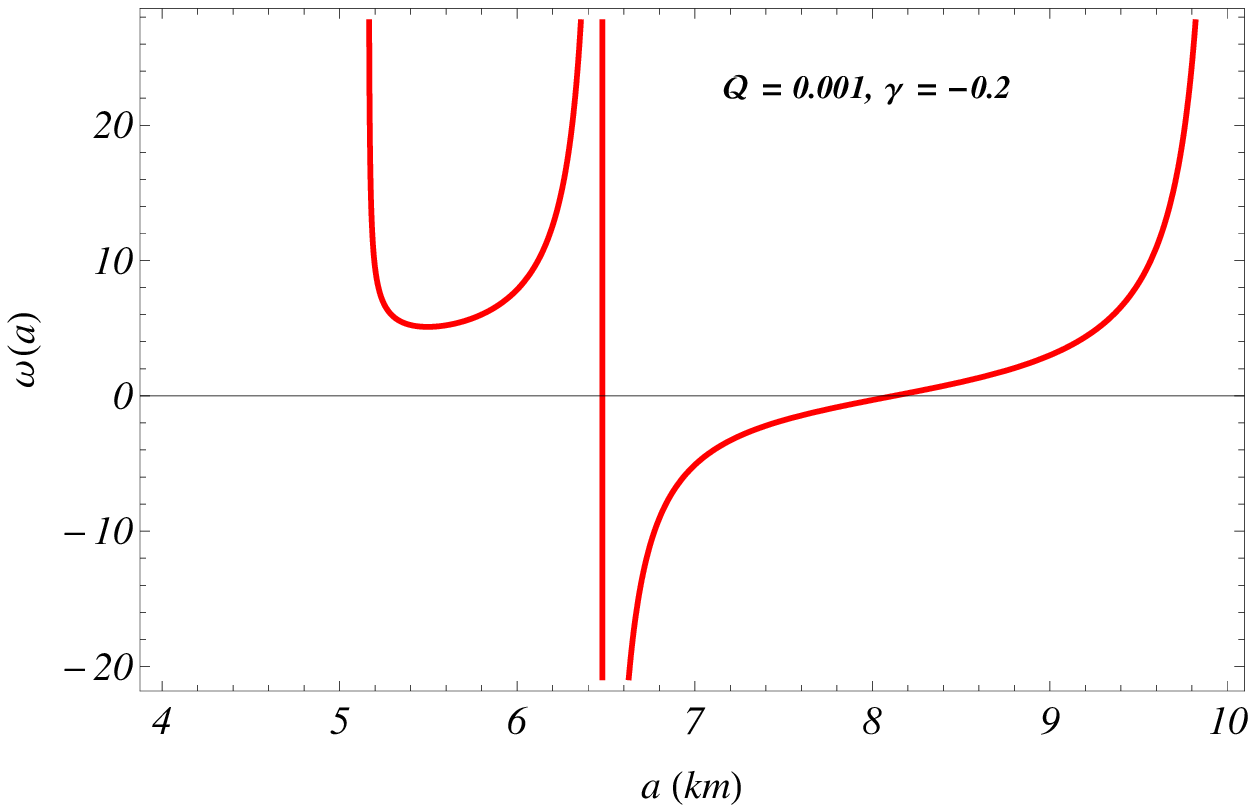}
        \includegraphics[scale=.5]{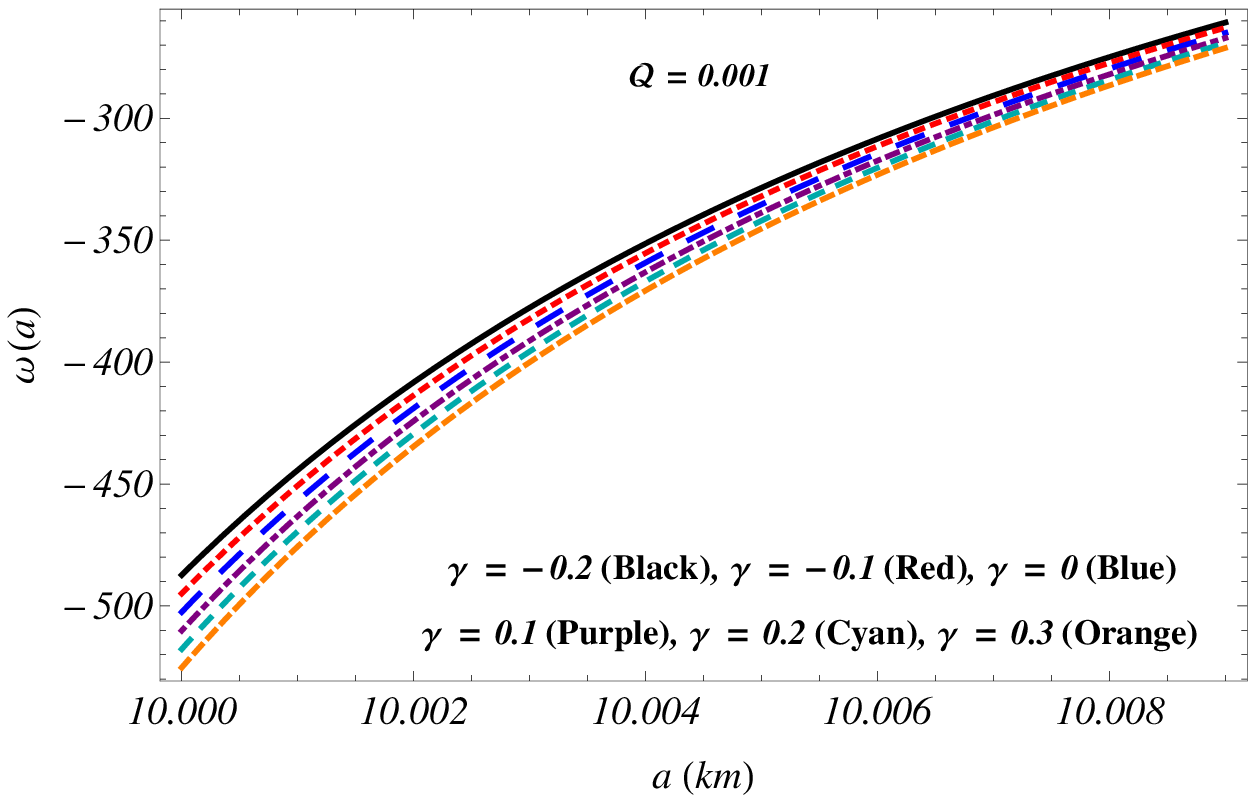}
        \includegraphics[scale=.5]{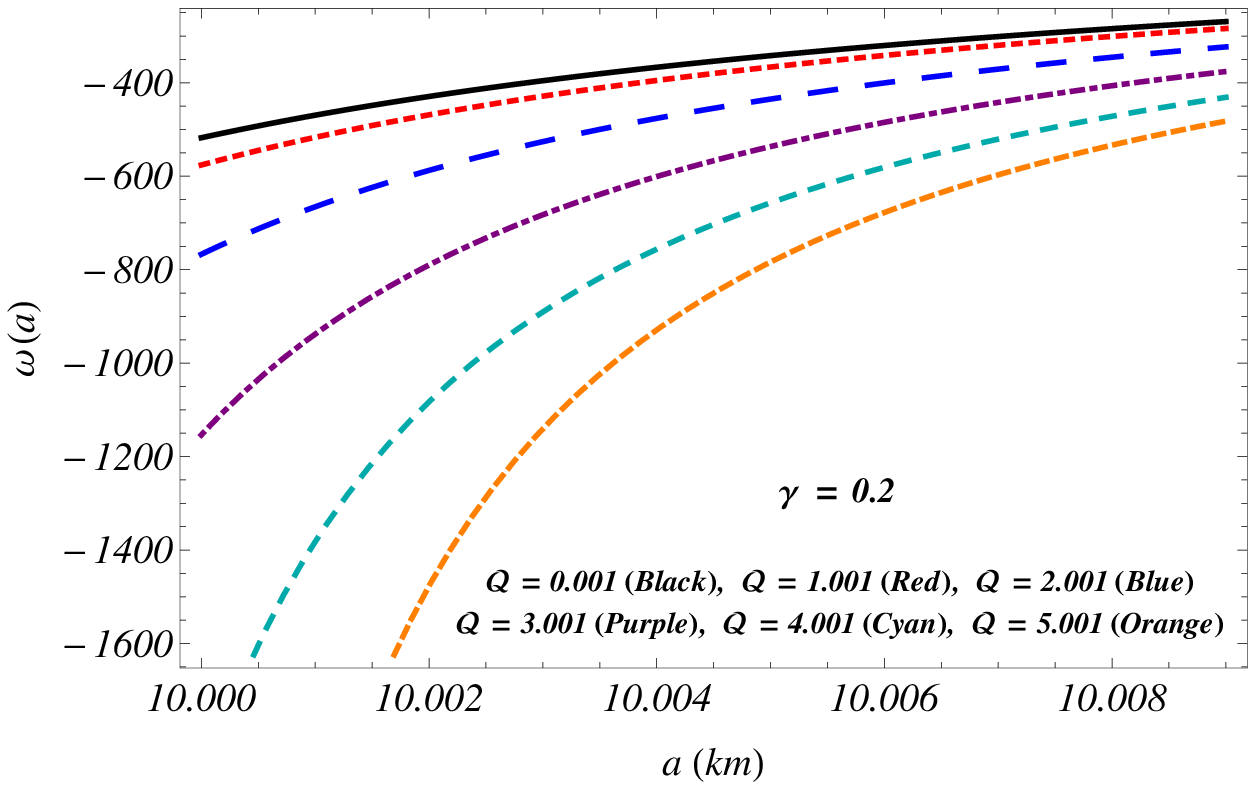}
       \caption{Top two profiles show variation of the equation of state parameter in the interior region, bottom two panels show variation of equation of state parameter inside the thin shell.\label{omega}}
\end{figure}
To keep $\omega(a)$ real, we need the restriction $\frac{2\mathcal{M}}{a}-\frac{\mathcal{Q}^2}{a^2}<1$, which is already satisfied from the eqn. (\ref{exterior}). Now $\omega(a)$ may be positive or negative depending on the signature of numerator or denominator of eqn. (\ref{omega1}). The profiles of $\omega(a)$ against `a' are shown in Fig.~\ref{omega}. The location of the thin shell (junction surface) plays an important role: if `a' is sufficiently large, then $\omega(a) \rightarrow -1$ and it incorporates the dark energy effects of the cosmological constant $\Lambda$. For very small value of `a' $\omega(a)$ tends to zero yielding a dust shell. According to Fig.~\ref{omega}, it can be observed that $\omega(a)$ is negative within the thin shell, implying that $\mathcal{P}(a)$ and $\sigma(a)$ are the opposite sign. The surface pressure is negative, which means a tension. In the junction shell, the energy density is positive. The thin shell, i.e. region II in our configuration, contains ultra-relativistic fluid obeying the relationship $p=\rho$ and the second fundamental form of discontinuity provides additional surface stress energy and surface tension for the connecting interface. These two non-interacting components are characteristic features of our non-vacuum region II.

\subsection{Proper length of the shell}
We assume the lower and upper boundaries of the shell are $r = a$ and
$r = a+\epsilon$ respectively and hence the proper thickness of the shell
is $\epsilon$ which is a very small positive real number
$(0~<\epsilon\ll 1)$. The proper length $\mathcal{L}$ of such a region that
connects inner and outer boundary, can be obtained as,
\begin{eqnarray}\label{l0}
   \mathcal{L}&=& \int_{a}^{a+\epsilon}\frac{1}{\sqrt{e^{-\lambda}}} dr=\int_{a}^{a+\epsilon}\frac{1}{\sqrt{\frac{2\gamma+\kappa}{2\kappa-4\gamma}-Dr^{\frac{2\gamma-\kappa}{4\gamma+\kappa}}}}dr,\nonumber\\
  &=&\bigg[r\frac{\kappa-2\gamma}{\kappa+2\gamma}\sqrt{\frac{2\kappa+4\gamma}{\kappa-2\gamma}-\frac{4D}{r^{\frac{\kappa-2\gamma}{\kappa+4\gamma}}}}\zeta(r)\bigg]_a^{a+\epsilon}.
\end{eqnarray}
\begin{figure}[htbp]
        \includegraphics[scale=.5]{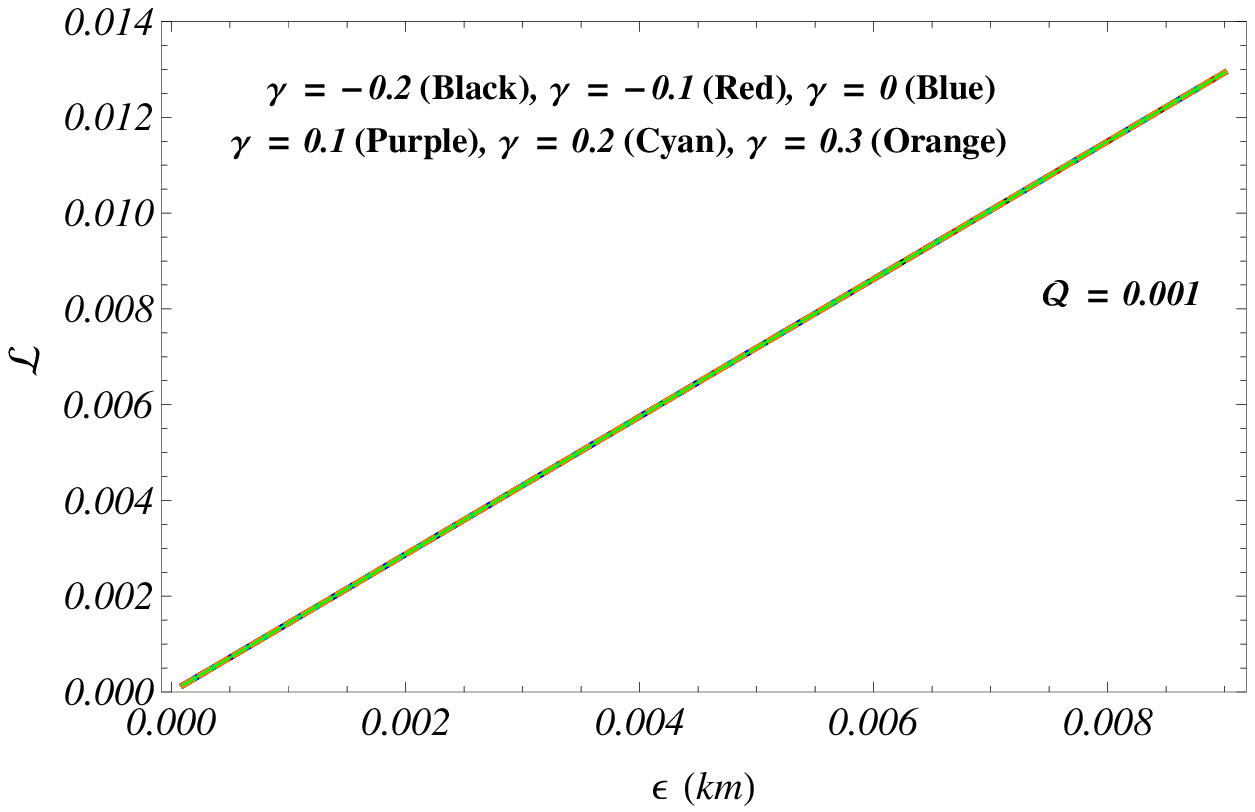}
        \includegraphics[scale=.5]{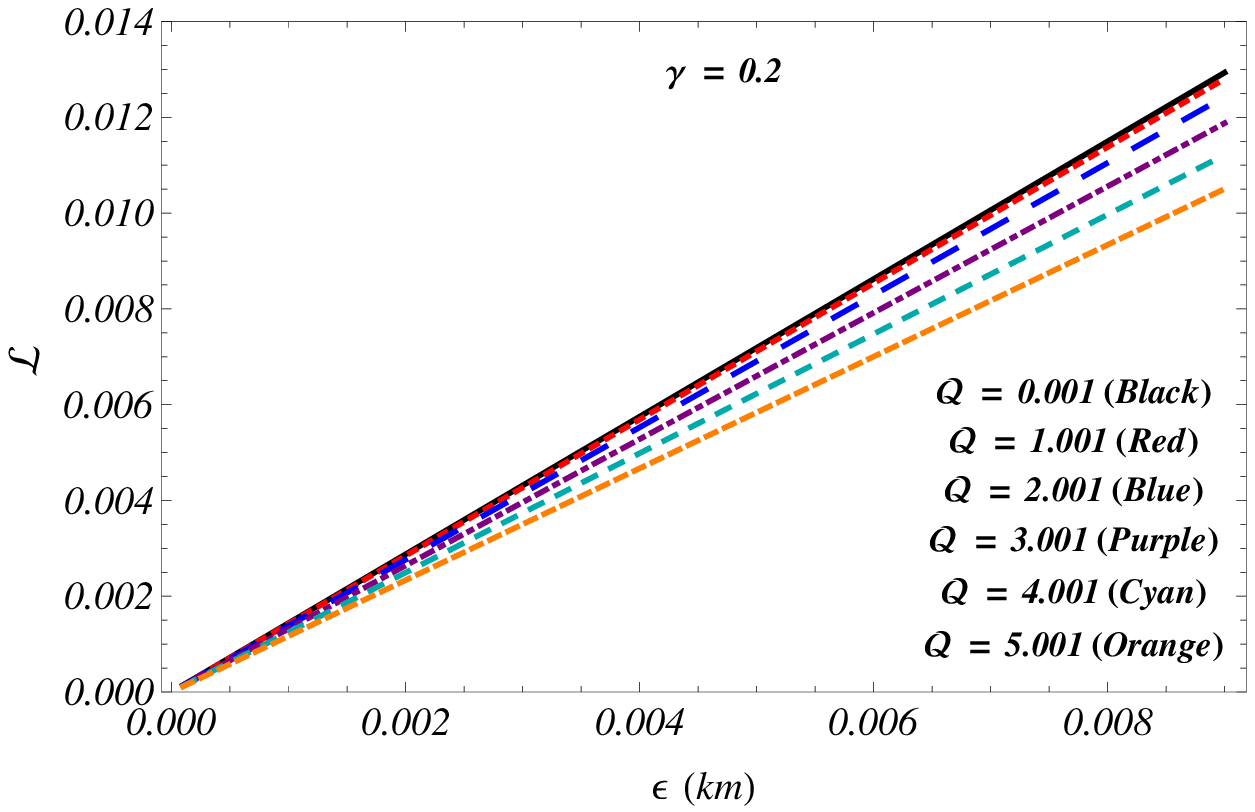}
       \caption{Variation of proper length inside the shell versus thickness of the shell
of charged gravastar. \label{ell}}
\end{figure}

where, \begin{eqnarray*}\zeta(r)&=&_2F_1\left[1, \frac{10 \gamma + \kappa}{
 4 \gamma - 2 \kappa}, -\frac{6 \gamma}{-2 \gamma + \kappa}, \frac{2D(\kappa-2\gamma)}{(\kappa+2\gamma)r^{\frac{\kappa-2\gamma}{\kappa+4\gamma}}}\right],\end{eqnarray*}
 and $_2F_1$ is the hypergeometric function defined as
\[_2F_1(a, b; c; x) = \sum_{k=0}^{\infty}\frac{(a)_k(b)_k}{(c)_k}\frac{x^k}{k!} \] and the series expansion of R.H.S of the above equation is,
\[1+\frac{a\cdot b}{c}\frac{x}{1!}+\frac{a(a + 1)b(b + 1)}{c(c + 1)
}\frac{x^2}{2!}+...\]
where, $|x| < 1$; $a,\, b,\,c$ are real numbers and $c \neq
0, -1, -2, . . .$ Here $(a)_n$ ($n$ is a positive integer) being
Pochhammer symbol defined by,
$(a)_n = a(a + 1)\cdot\cdot\cdot(a + n - 1),$
with $(a)_0 = 1$.
With the help of simple algebra, $(a)_n$ takes the form,
$(a)_n =\frac{\Gamma(a+n)}{\Gamma(a)}$. The graphical behavior of the proper length with respect to the thickness
of thin-shell gravastar is displayed in Fig.~\ref{ell}
\subsection{Entropy}
Entropy is used to the measure of disorderness or disturbance in a mechanical system. According to the
theory of Mazur and Mottola \cite{mazur2001,mazur2004}, charged gravastar has zero entropy density for
the interior region. Using the concept of Mazur and M\"{o}ttola \cite{mazur2001,mazur2004}, the entropy within the shell of the the charged gravastar is calculated as,
\begin{equation}\label{l4}
S=\int_{a}^{a+\epsilon}\frac{\kappa}{2} r^2 h(r)\sqrt{e^{\lambda}}dr,
\end{equation}
By the standard thermodynamic
relation, $Ts = p + \rho$ for a relativistic fluid with zero
chemical potential and at the local temperature $T(r)$, the entropy density $s(r)$ can be expressed
as $$s(r)=\frac{2\alpha^2 K_B^2 T(r)}{\kappa\hbar^2}=2\alpha \left(\frac{K_B}{\hbar}\right)\sqrt{\frac{p(r)}{\kappa}}$$ Where $\alpha$ is a dimensionless constant, $K_B$ representing the Boltzmann constant, $\hbar=\frac{h}{2\pi}$, where $h$ is the Planck constant.
Using the expression for $p$ and $e^{\lambda}$, from eqn.(\ref{l4}), we calculate the expression for entropy as follows:
\begin{widetext}
\begin{eqnarray}\label{l5}
S&=&\int_a^{a+\epsilon}\frac{\kappa \alpha}{2\sqrt{4\gamma+\kappa}} \left(\frac{K_B}{\hbar}\right)r \sqrt{\frac{3 D (2 \gamma - \kappa) r^{\frac{2 \gamma}{
  4 \gamma + \kappa}} + (4 \gamma + \kappa) r^{\frac{\kappa}{4 \gamma + \kappa}}}{2 D (2 \gamma - \kappa) r^{\frac{2 \gamma}{
  4 \gamma + \kappa}} + (2 \gamma + \kappa) r^{\frac{\kappa}{4 \gamma + \kappa}}}} dr,
  \end{eqnarray}
\end{widetext}
  The above eqn. (\ref{l5}) can be written as, \[S=\frac{\kappa \alpha}{2\sqrt{4\gamma+\kappa}} \left(\frac{K_B}{\hbar}\right) \mathcal{I},\]
  where,
\begin{equation}\label{l}
\mathcal{I}=\int_{a}^{a+\epsilon}G(r)dr,
\end{equation}

and $G(r)$ is function of `r' defined as, \[G(r)=\sqrt{\frac{3 D (2 \gamma - \kappa) r^{\frac{2 \gamma}{
  4 \gamma + \kappa}} + (4 \gamma + \kappa) r^{\frac{\kappa}{4 \gamma + \kappa}}}{2 D (2 \gamma - \kappa) r^{\frac{2 \gamma}{
  4 \gamma + \kappa}} + (2 \gamma + \kappa) r^{\frac{\kappa}{4 \gamma + \kappa}}}}.\]

  \begin{figure}[htbp]
        \includegraphics[scale=.5]{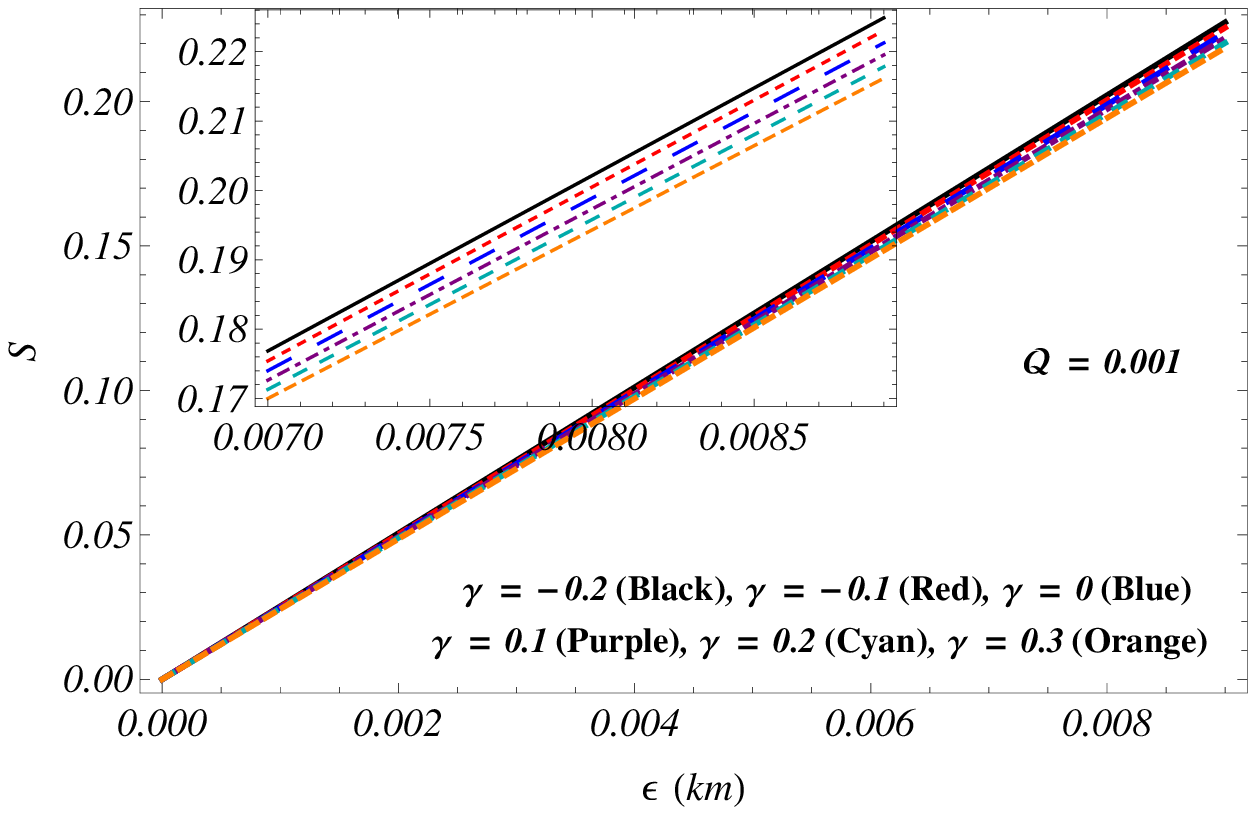}
         \includegraphics[scale=.5]{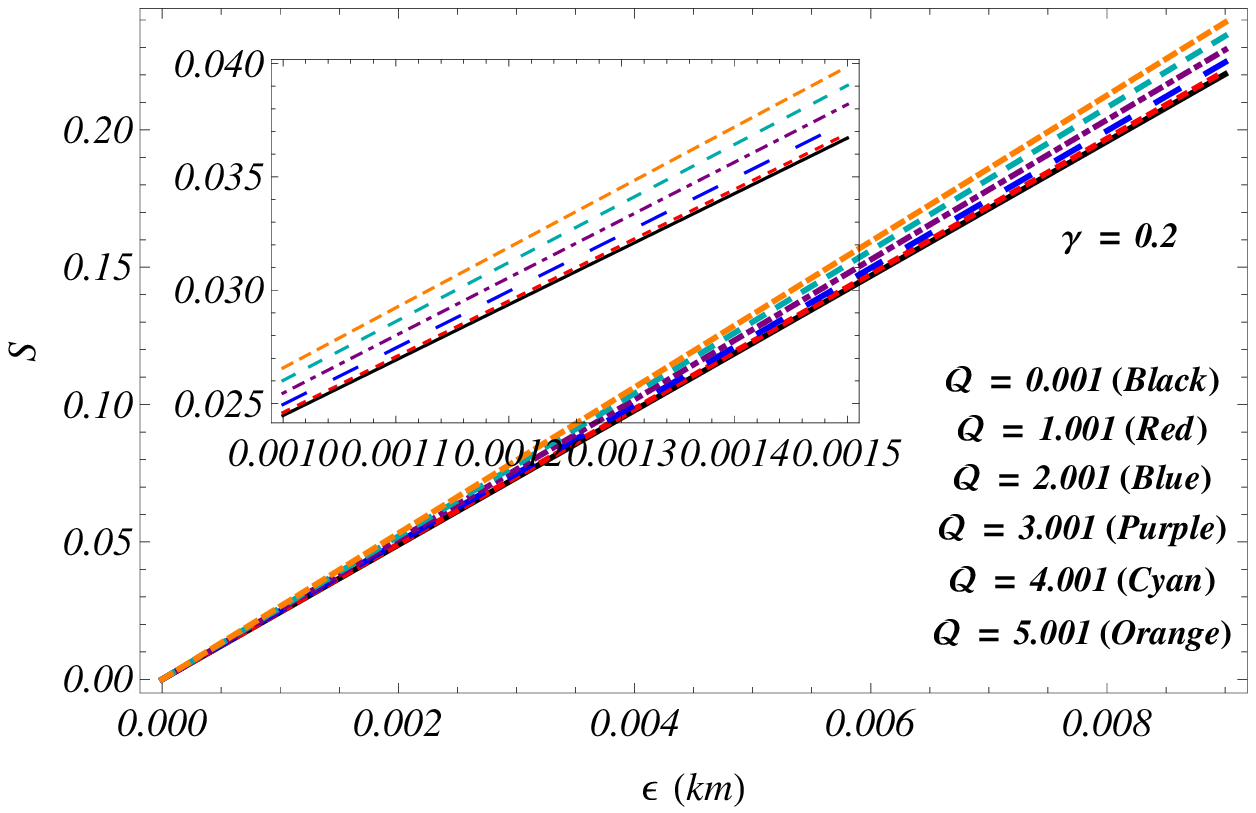}
       \caption{Variation of entropy inside the shell versus thickness of the shell
of charged gravastar. \label{entropy}}
\end{figure}

  Now due to the complexity of the expression of $G(r)$, it is very difficult to perform the integral given in eqn.(\ref{l}). let $H(r)$ be function such that $\frac{dH(r)}{dr}=G(r)$. Then, from equation (\ref{l}), by using the fundamental theorem of integral calculus, we obtain,
\begin{equation}\label{l1}
\mathcal{I}=\left[H(r)\right]_{a}^{a+\epsilon}=H(a+\epsilon)-H(a),
\end{equation}
Now by considering the Taylor series expansion of $H(a+\epsilon)$ about `a' and retaining up to the linear order of $\epsilon$, we obtain
\begin{equation}\label{l2}
\mathcal{I}=a\epsilon \sqrt{\frac{3 D (2 \gamma - \kappa) a^{\frac{2 \gamma}{
  4 \gamma + \kappa}} + (4 \gamma + \kappa) a^{\frac{\kappa}{4 \gamma + \kappa}}}{2 D (2 \gamma - \kappa) a^{\frac{2 \gamma}{
  4 \gamma + \kappa}} + (2 \gamma + \kappa) a^{\frac{\kappa}{4 \gamma + \kappa}}}},
\end{equation}
and consequently from equation (\ref{l5}), we get,
\begin{widetext}
  \begin{eqnarray}\label{ent}
S&=&\frac{\kappa \alpha}{2\sqrt{4\gamma+\kappa}} \left(\frac{K_B}{\hbar}\right)a\epsilon \sqrt{\frac{3 D (2 \gamma - \kappa) a^{\frac{2 \gamma}{
  4 \gamma + \kappa}} + (4 \gamma + \kappa) a^{\frac{\kappa}{4 \gamma + \kappa}}}{2 D (2 \gamma - \kappa) a^{\frac{2 \gamma}{
  4 \gamma + \kappa}} + (2 \gamma + \kappa) a^{\frac{\kappa}{4 \gamma + \kappa}}}}.
\end{eqnarray}
\end{widetext}
Hence we have successfully obtained the expression of the entropy for our proposed model. From eqn. (\ref{ent}) one can note that if the thickness of the thin shell $\epsilon\ll a$, then $S\approx \mathcal{O}(\epsilon)$. In ref. \cite{usmani2011}, Usmani et al. showed that the entropy
depends on the thickness of the shell. Our result is consistent with the result of ref. \cite{usmani2011}. The variation of entropy with respect to the thin shell radius is shown in Fig.~\ref{entropy}.
\subsection{Energy within the thin shell}
Let us now calculate the energy $\mathcal{E}$ within the shell from the following formula
\begin{eqnarray}\label{l3}
\mathcal{E}&=&\int_{a}^{a+\epsilon}\frac{\kappa}{2} r^2\left(\rho+\frac{E^2}{\kappa}\right)dr\nonumber\\
&=&\left[\frac{(4\gamma - \kappa) r}{4 (2\gamma -\kappa)}+\frac{D}{4}r^{\frac{6\gamma}{4\gamma+\kappa}}\right]_a^{a+\epsilon},\nonumber\\
&=&\frac{\epsilon}{4}\left(\frac{4\gamma - \kappa}{2\gamma -\kappa}\right)+\frac{D}{4}\left[(a+\epsilon)^{\frac{6\gamma}{4\gamma+\kappa}}-a^{\frac{6\gamma}{4\gamma+\kappa}}\right],\nonumber\\
&\approx& \frac{\epsilon}{4}\left[\frac{4\gamma - \kappa}{2\gamma -\kappa}+\frac{6D\gamma}{4\gamma+\kappa} \right].
\end{eqnarray}

\begin{figure}[htbp]
         \includegraphics[scale=.5]{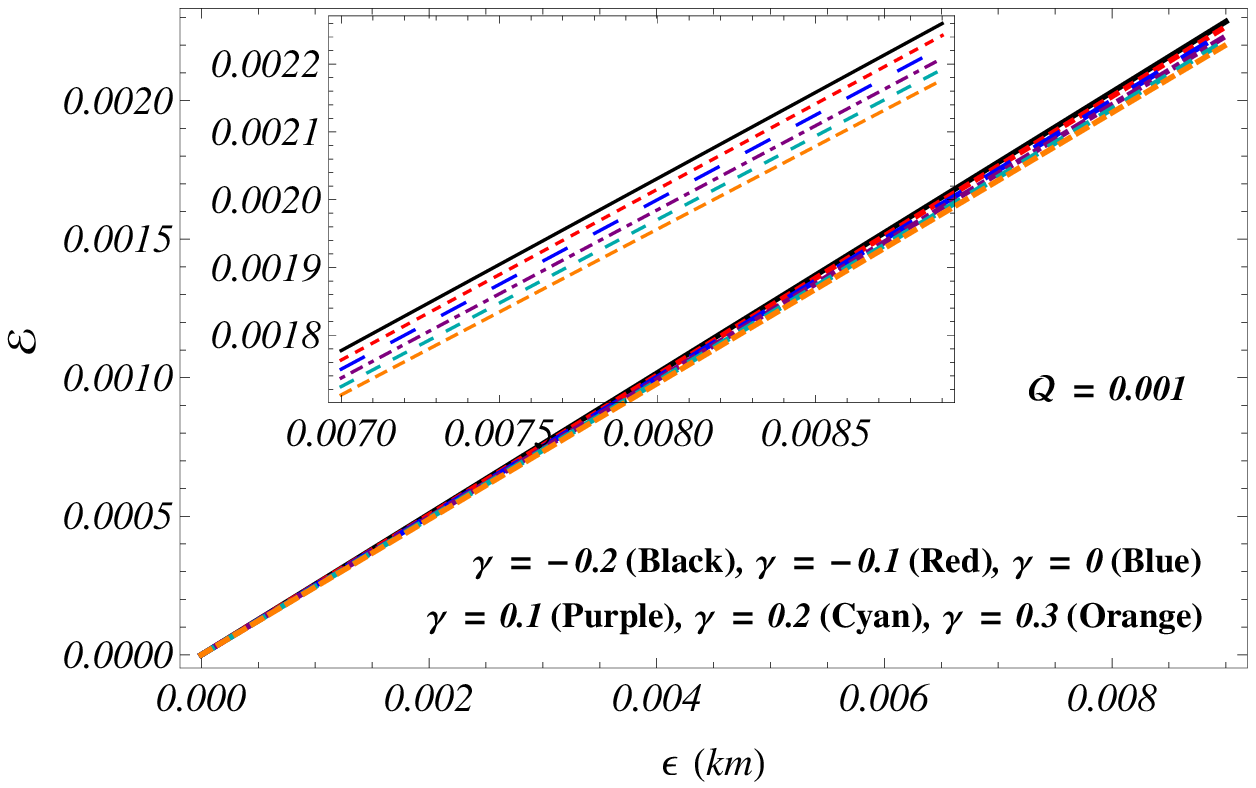}
          \includegraphics[scale=.5]{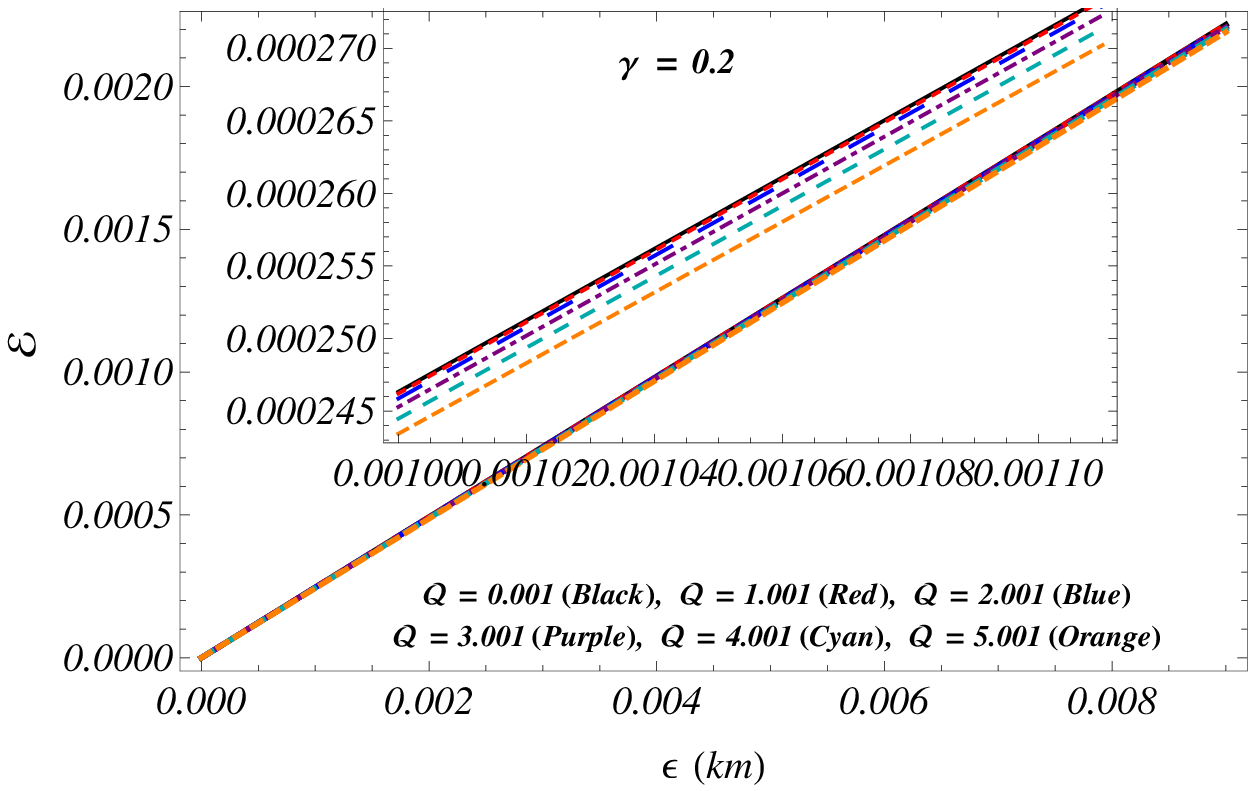}
       \caption{Variation of energy inside the shell versus thickness of the shell
of charged gravastar. \label{energy}}
\end{figure}

This shows a direct relation of the energy with the thickness
of the shell. From expression (\ref{l3}) we see that, the energy is directly proportional to the thickness
of the shell, so unit of energy is also `km'. The graphical analysis of energy-thickness relation corresponding to different values of $\gamma$ is given in Fig. \ref{energy}
which displays the non-repulsive nature of energy inside the
shell.

\section{Stability of the gravastar}\label{stability1}
In this section, we are interested to check the stability of gravastars. For this purpose  we define a new parameter $\eta$ as the ratio of the derivatives of $\sigma$ and  $\mathcal{P}$ as follows,
\begin{eqnarray}
\eta(a)=\frac{\mathcal{P'}(a)}{\sigma'(a)}.
\end{eqnarray}

\begin{figure}[htbp]
        \includegraphics[scale=.5]{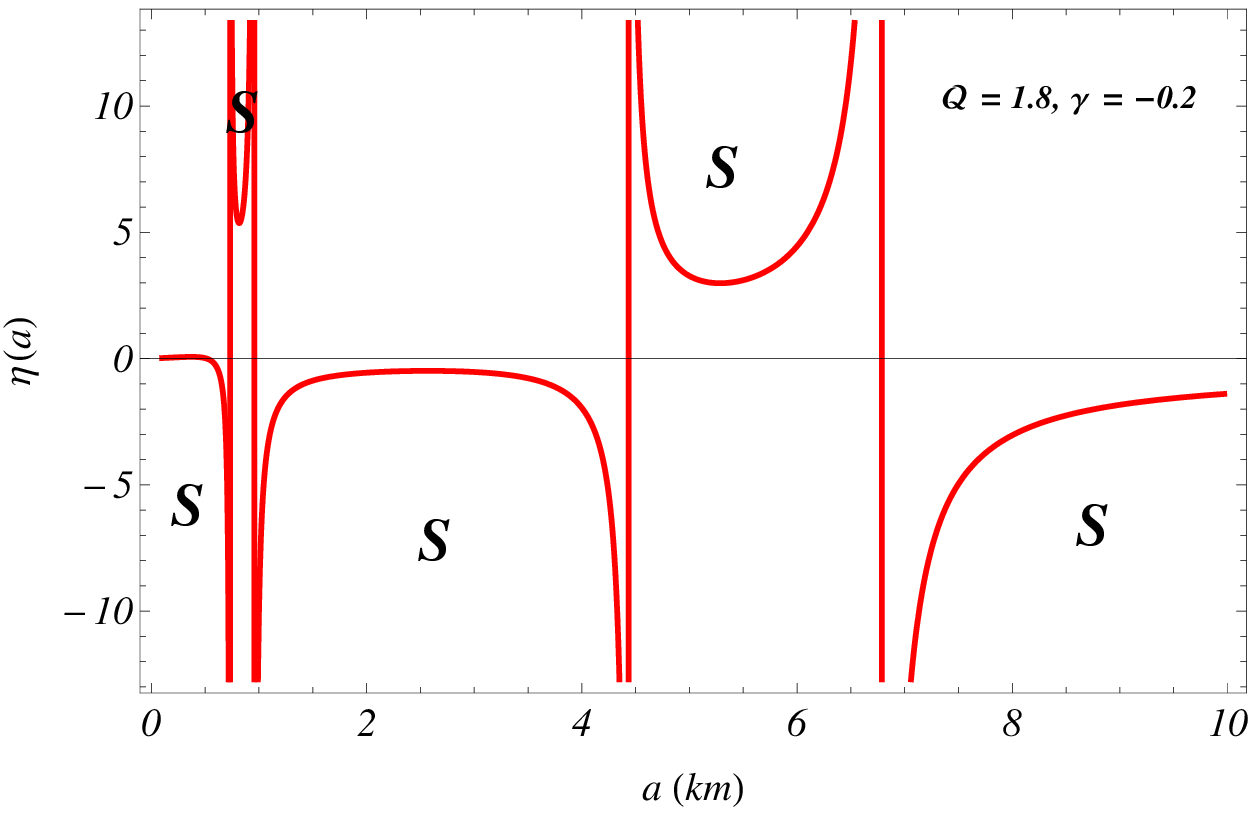}
        \includegraphics[scale=.5]{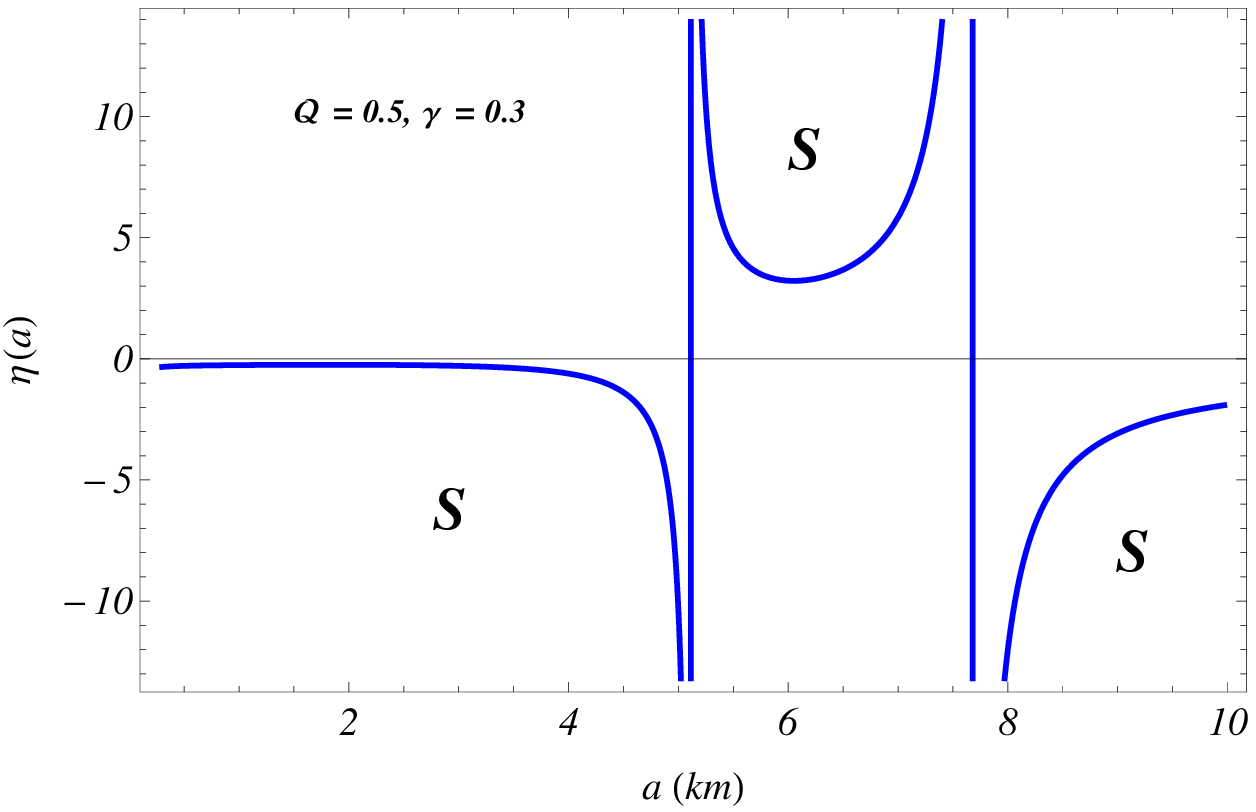}
        \includegraphics[scale=.5]{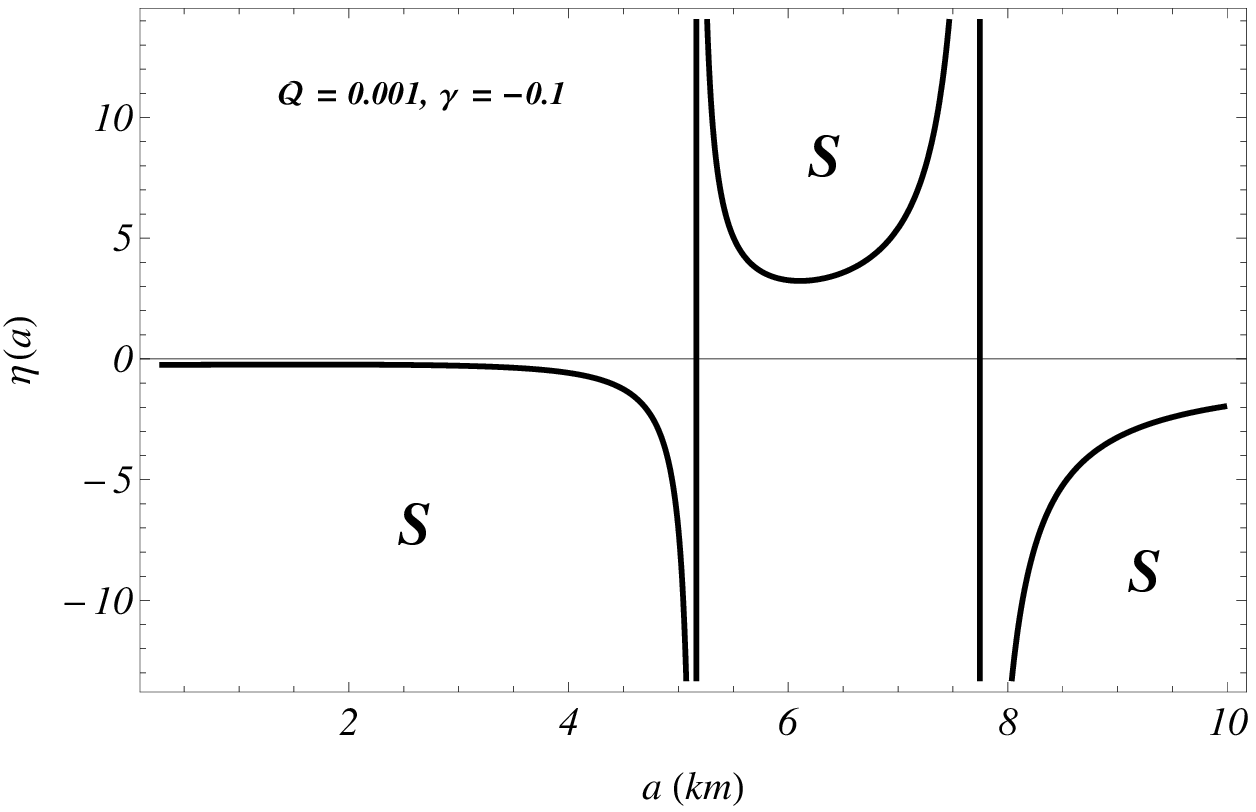}
       \caption{Stability regions of the charged gravastar. Here `S' stands for stability. \label{eta}}
\end{figure}

\"{O}vg\"{u}n et al.\cite{Ovgun:2017jzt} used the above parameter which plays a fundamental role in
determining the stability regions of the respective solutions of charged thin-shell
gravastar model within the context of noncommutative geometry.
Yousaf et al. \cite{yousaf2019} also discussed the stability of the gravastar model in $f(R,\,T)$ gravity in presence of charge. Debnath \cite{Debnath:2019eor} obtained the stable regions of the charged gravastar in Rastall-Rainbow gravity. Very recently Sharif and Javed \cite{Sharif:2021zzr} studied the stability of thin-shell gravastars and conclude that stable regions of gravastar shell decrease and dynamical configuration increase with cosmological constant. The parameter $\eta$ is interpreted as the squared speed of
sound and it should satisfy $0\leq \eta \leq 1$ since the speed of sound
should not exceed the speed of light. But according to Poisson and Visser \cite{Poisson:1995sv}, Lobo \cite{Lobo:2003xd} the range of $\eta$ may
be lying outside the range mentioned earlier on the surface layer. For our present study, we have plotted the profile of $\eta$ in Fig.~\ref{eta} for different values of $\mathcal{Q}$ (keeping $\gamma$ fixed) and different values of coupling constant $\gamma$ (keeping $\mathcal{Q}$ fixed) mentioned in the figures and the stability region have been identified. The details calculations regarding stability analysis for gravastar can be found in \cite{Ovgun:2017jzt,Sharif:2021zzr} and the similar type of stability regions were obtained in the refs.\cite{yousaf2019,Ovgun:2017jzt,Sharif:2021zzr}. It confirms the physical validity of our present model.
\section{Discussions and final remarks}\label{sec5}
In the present article we have studied the effects of modified gravity on charged gravastar corresponding to the exterior Reissner-Nordstr\"{o}m line element. In this section we are summarizing some key physical features of the model as follows. Several physical parameters, e.g. metric potentials, proper length of the shell, entropy, equation of state (EoS), energy within shell, surface redshift etc. have been discussed through both analytically and graphically. We have drawn all the physical parameters in Figs. \ref{nu}-\ref{eta}. The figures indicate the physical acceptability of our present model of charged gravastar. The metric coefficient $e^{\nu}$ is plotted against $r$ inside the thin shell is shown in Fig. \ref{nu}. $e^{\nu}$ does not depend on $\gamma$ rather it relies on $\mathcal{Q}$. To draw the profiles we have varied $\mathcal{Q}$ and note that $e^{\nu}$ takes higher value with increasing value of $\mathcal{Q}$. The another metric coefficient $e^{-\lambda}$ is depicted in Fig. \ref{lambda}. The value of $e^{-\lambda}$ at the inner boundary of the thin shell decreases as $\gamma$ increases when $\mathcal{Q}$ is fixed but at the outer boundary of the thin shell all the profiles of $e^{-\lambda}$ for different values of $\gamma$ coincide (keeping $\mathcal{Q}$ fixed). On the other hand, when $\gamma$ is fixed, $e^{-\lambda}$ takes higher values with increasing values of $\mathcal{Q}$ inside the thin shell. The nature of pressure (=density) inside the thin shell is depicted in Fig. \ref{pr}. The figure indicates that the pressure almost maintain the linear behaviour with the thickness of the thin shell. Moreover, within the thin shell the profile of $p$ takes lower value for increasing value of $\gamma$ when $\mathcal{Q}$ is fixed. The reverse nature of $p$ is seen for increasing value of $\mathcal{Q}$ when $\gamma$ is fixed. The surface energy density has been calculated
by following the condition of Darmois and Israel \cite{Darmois1927,Israel:1966rt}. The surface energy density
has been plotted against the radial parameter in Fig. \ref{se}.
The surface energy density remained positive throughout the shell and it gradually increases as we moved from
the inner boundary of the thin shell to the outer boundary. The value of the surface energy density at the inner boundary decreases with the increasing value of the coupling constant $\gamma$ when $\mathcal{Q}$ is fixed. Again for a fixed value of $\gamma$, the value of $\sigma$ at the interior boundary decreases as $\mathcal{Q}$ increases but its nature changes at the outer boundary. Fig. \ref{sp} shows the behaviour of surface pressure inside the thin shell and it can be seen that the surface pressure is negative in the interior of the shell. The values of $\mathcal{P}$ decreases with the increasing values of $\mathcal{Q}$ when $\gamma$ is fixed. On the other hand, values of $\mathcal{P}$ increases with the increasing values of $\gamma$ when $\mathcal{Q}$ is fixed. The equation of parameter, which is the ratio of the surface pressure and surface energy density, is depicted in Fig.~\ref{omega}. From the figure it is clear that $\omega(a)$ takes both positive and negative values inside the interior of the gravastar. Inside the thin shell region $\omega(a)$ takes negative values. It can be noted that inside the thin shell $\omega(a)$ decreases with the increasing values of $\gamma$ when $\mathcal{Q}$ is fixed. The same behaviour of $\omega(a)$ is noticed when $\mathcal{Q}$ varying with a fixed $\gamma$. From the profile given in Fig.~\ref{ell} of proper length of thin shell, it is seen that the coupling constant $\gamma$ has negligible effect on $\mathcal{L}$ for a fixed values of $\mathcal{Q}$ since all the profiles of $\mathcal{L}$ coincides for different values of $\gamma$. On contrary, $\mathcal{L}$ takes lower values when $\mathcal{Q}$ increases for fixed values of $\gamma$. In both cases $\mathcal{L}$ takes positive values inside the shell. The profile of entropy with respect to the thickness of the thin shell is depicted in Fig.~\ref{entropy}. The entropy is a increasing function of $\epsilon$, it takes the maximum values at the outer boundary of the thin shell. For fixed values of $\mathcal{Q}$, entropy decreases with the increasing values of $\gamma$, reverse behaviour is noticed when $\mathcal{Q}$ varies. We have also plotted the energy within the thin shell with respect to the thickness of the thin shell in Fig.~\ref{energy}. The profile of energy within the thin shell takes lower values with the increasing values of $\gamma$ when $\mathcal{Q}$ is fixed. The same nature of energy within the thin shell can be noticed
for different values of $\mathcal{Q}$ when $\gamma$ is fixed. The stability analysis of the present model is discussed and stable regions are marked in Fig. \ref{eta}. \par
Now we want to discuss about the possible observational signatures for this kind of gravastar. Till now there are no direct evidences to detect garavastar but few indirect ways are available in the literature which gives clues of their possible existence and future detection of gravastar. We can adopt a spherical thin-shell gravastar model that links interior de Sitter geometry with exterior Schwarzschild geometry \cite{visser2004}.
Sakai et al. \cite{sakai} first proposed the concept for possible detection mechanism of gravastar through the study of  gravastar shadows. Gravitational lensing effects \cite{kubo} can be used as another method to identify gravastar where they proposed that in a gravastar microlensing effects with a higher maximal luminosity than black holes of the same mass might occur. To detect gravastar, they presented the following two models:
\begin{enumerate}[I.]
  \item According to Model 1, they calculate the image of a companion rotating around the gravastar and discover that certain characteristic images arise depending on whether the gravastar has photon orbits that are unstable or not (assuming the surface of thin-shell gravastar to be optically transparent)
  \item According to Model 2, they compute the microlensing effects, the overall luminosity change, and the peak luminosity may be far greater than a black hole of the same mass.
  \end{enumerate}
It should be mentioned that interferometric LIGO detectors have just detected the ringdown signal of GW 150914 \cite{abbott2016}. Observational constraints on gravastar models with the thermal process was studied by Broderick and Narayan \cite{Broderick}. Uchikata et al. \cite{Uchikata} proposed that to constrain gravastars, measurement of the tidal deformability from the gravitational-wave detection of a compact-binary inspiral can be used. To rule out exotic alternatives to BHs and to test quantum effects at the horizon scale, only late-time ringdown detections might be used \cite{car1,car2}. They anticipated that objects with no event horizon, such as gravastars, would be the source of such Gravitational Waves with a high probability. According to Chirenti and Rezzolla \cite{Chirenti2016}, we cannot assert that gravastars merging is the source of Gravitational Waves because we know so little about the perturbative reaction of rotating gravastars. Recently, after analysing the image acquired in the First M87 Event Horizon Telescope (EHT) \cite{EHT2019} finding, it was discovered that the generated shadow might be attributable to gravastar. Shadow may be cast by any compact object with a spacetime defined by unstable circular photon orbits, as demonstrated by Mizuno et al.\cite{Mizuno2018}.

The gravastar theory may also be examined in the framework of Friedmann's flat universe cosmology \cite{Croker2016}. This study is motivated by the action principle itself, and the conclusion is quite fascinating, in which the gravastar population produces a dynamic kind of dark energy. If comparable effects to those stated above are observed in the future, it will provide an excellent foundation for comparing GR with modified gravity. \par
From all the presented results for our present study, we can conclude that, the $f(R,\,T)$ gravity leads to very distinct
gravastar model in presence of charge even when the spacetime admits a conformal killing vector. Our results agree with the results obtained by Usmani et al. \cite{usmani2011} in the case $\gamma \rightarrow 0$. In the $f(R,\,T)$ theory of gravity, we derive solutions that adequately explain gravastars. After careful consideration, we conclude that our gravastar model is stable under $f(R,\,T)$ theory of gravity, which differs conceptually from Einstein's GR. As a final comment, we can state that there are no direct observable evidences that can distinguish a black hole from a gravastar at this time. The new findings of GW190521 once again demonstrate that the black hole hypothesis is incompatible with observable results. As a result, it's possible that the hypothetical black hole is a gravastar.

\section*{Acknowledgments}
 P.B. is thankful to the Inter University Centre for Astronomy and Astrophysics (IUCAA), Government of India, for providing visiting associateship.

\bibliography{epjc3}

\end{document}